\documentclass{article}

\PassOptionsToPackage{numbers,comma,sort,compress}{natbib}
\usepackage[preprint]{neurips_2026}

\usepackage[table,dvipsnames]{xcolor}     
\usepackage{pifont}
\usepackage{graphicx}
\usepackage{subcaption}
\usepackage{graphicx}
\usepackage{multirow} 
\usepackage{amsmath}
\usepackage{adjustbox}
\usepackage{arydshln}
\usepackage{float}
\usepackage{amssymb}
\usepackage{xspace}
\usepackage{makecell}
\usepackage{tabu}
\usepackage{fontawesome5}
\usepackage{wrapfig}
\usepackage{comment}
\usepackage{arydshln}
\usepackage{algorithm,algorithmic}

\usepackage{enumitem}
\setlist[enumerate]{nosep,leftmargin=*}

\usepackage{colortbl}
\usepackage{tabularx}
\usepackage{booktabs}
\usepackage{ragged2e}
\usepackage{soul}
\usepackage{bbm}

\sethlcolor{gray!15}
\soulregister\cite7
\soulregister\ref7
\soulregister\eqref7

\newlist{tinenum}{enumerate}{1}
\setlist[tinenum]{
  label=\arabic*., leftmargin=1.5em,
  topsep=0pt, partopsep=0pt, itemsep=0pt, parsep=0pt,
  before=\vspace{-0.7\baselineskip},
  after=\vspace{-1\baselineskip}
}

\newcommand{\ourmethod}{AdaM-Rec\xspace}

\newcommand{\ourtitle}{
AdaM-Rec: Adaptive Modality Routing for Multimodal Recommendation
}

\expandafter\def\expandafter\normalsize\expandafter{%
    \normalsize%
    \setlength\abovedisplayskip{2pt}%
    \setlength\belowdisplayskip{2pt}%
    \setlength\abovedisplayshortskip{-8pt}%
    \setlength\belowdisplayshortskip{2pt}%
}

\definecolor{navyblue}{HTML}{0071BC}
\definecolor{hotpink}{HTML}{FF0080}
\definecolor{oai-white}{HTML}{FFFFFF}
\definecolor{oai-black}{HTML}{000000}
\definecolor{oai-red}{HTML}{FF4500}
\definecolor{oai-green}{HTML}{51DA4C}
\definecolor{oai-blue}{HTML}{0000FF}
\definecolor{oai-yellow}{HTML}{FFF639}
\definecolor{oai-magenta}{HTML}{FF45FF}
\definecolor{oai-cyan}{HTML}{00FFFF}
\definecolor{oai-orange}{HTML}{FE7600}
\definecolor{oai-violet}{HTML}{8A2BE2}
\definecolor{oai-brown}{HTML}{A0522D}
\definecolor{oai-green-050}{HTML}{F4FFF4}
\definecolor{oai-green-100}{HTML}{E9FFE8}
\definecolor{oai-green-200}{HTML}{D9FFD8}
\definecolor{oai-green-300}{HTML}{C9FFC7}
\definecolor{oai-green-400}{HTML}{A6FFA3}
\definecolor{oai-green-500}{HTML}{7CF178}
\definecolor{oai-green-600}{HTML}{51DA4C}
\definecolor{oai-green-700}{HTML}{3FA93B}
\definecolor{oai-green-800}{HTML}{2D712A}
\definecolor{oai-green-900}{HTML}{193718}
\definecolor{oai-gray-000}{HTML}{FFFFFF}
\definecolor{oai-gray-100}{HTML}{FAFAFA}
\definecolor{oai-gray-200}{HTML}{F5F5F5}
\definecolor{oai-gray-300}{HTML}{E5E5E5}
\definecolor{oai-gray-400}{HTML}{FFB7A4}
\definecolor{oai-gray-500}{HTML}{CDCDCD}
\definecolor{oai-gray-600}{HTML}{A8A8A8}
\definecolor{oai-gray-700}{HTML}{747474}
\definecolor{oai-gray-800}{HTML}{393939}
\definecolor{oai-gray-900}{HTML}{000000}
\definecolor{Green}{rgb}{0.0, 0.5, 0.0}
\definecolor{Red}{rgb}{1.0, 0.0, 0.0}

\newcommand{\groupheader}[1]{%
  \multicolumn{6}{>{\columncolor[HTML]{FFF0E0}}l}{%
    \hspace{-0.35em}\textbf{\textit{\footnotesize #1}}%
  }\\%
}

\usepackage[utf8]{inputenc} 
\usepackage[T1]{fontenc}    
\usepackage{hyperref}       
\usepackage{url}            
\usepackage{booktabs}       
\usepackage{amsfonts}       
\usepackage{nicefrac}       
\usepackage{microtype}      

\usepackage{times}
\usepackage{latexsym}
\usepackage{setspace}

\usepackage[T1]{fontenc}

\usepackage[utf8]{inputenc}

\usepackage{microtype}

\usepackage{inconsolata}

\usepackage{graphicx}

\usepackage{tcolorbox}
\tcbuselibrary{skins,breakable}

\hypersetup{
colorlinks,
linkcolor={oai-red!50!oai-black},
citecolor={navyblue!50!black},
urlcolor={blue!80!black}
}

\usepackage{tcolorbox}   
\tcbuselibrary{skins}    
\tcbuselibrary{breakable}
\usepackage{tikz}        
\usepackage{varwidth}    
\usepackage{float} 

\usepackage[utf8]{inputenc}   
\usepackage[T1]{fontenc}      
\usepackage{times}             
\usepackage{inconsolata}       
\usepackage{microtype}         

\usepackage{amsmath, amssymb, amsfonts}
\usepackage{nicefrac}          

\usepackage{colortbl}          
\usepackage{booktabs}          
\usepackage{multirow}          
\usepackage{makecell}          
\usepackage{array}             
\usepackage{adjustbox}         

\usepackage{graphicx}          
\usepackage{tikz}              
\usepackage{subcaption}        
\usepackage{float,wrapfig}     

\usepackage{hyperref}          
\usepackage{url}               

\usepackage{algorithm}
\usepackage{algorithmic}

\usepackage{pifont}            
\usepackage{xspace}            
\usepackage{amsthm}            
\usepackage{soul}              
\usepackage{caption}           
\usepackage{balance}           

\usepackage{pifont} 

\definecolor{cgreen}{RGB}{34,139,34} 
\definecolor{cred}{RGB}{178, 34, 34}

\newcommand{\tightbottomrule}{\specialrule{\heavyrulewidth}{0pt}{0pt}}

\title{\ourtitle}

\author{Honghao Fu$^{1, 2}$, Jiacheng Chen$^{2}$, Manxi Lin$^{2}$, Junjun Zheng$^{2}$, Xiangheng Kong$^{2}$\\
\textbf{Yiwei Wang$^{3}$, Xin Yu$^{4}$, Miao Xu$^{1}$, Yuning Jiang$^{2}$, Yujun Cai$^{3}$\thanks{Corresponding Author}} \\
$^1$University of Queensland, \; $^2$Alibaba Group, \; $^3$Southeast University, \; $^4$Adelaide University\\
}

\begin{document}

\maketitle

\begin{abstract}
While recent multimodal recommender systems have demonstrated the effectiveness of incorporating visual and textual information to improve downstream performance, most existing methods rely on static modality fusion, assuming that the relative importance of textual and visual signals remains stable across recommendation scenarios. This design may not fully account for an important variation across recommendation requests: some queries require fine-grained visual cues, whereas others are better served by textual or functional semantics, in which case indiscriminate modality fusion brings in uninformative cues and impairs recommendation quality. To address this, we propose {\bf \ourmethod}, an LLM-based framework for adaptive modality routing in multimodal recommendation, which enables dynamic calibration of reliance on textual and multimodal evidence for user-specific queries. Built on structured natural-language representations of items and user preferences, it estimates modality reliability using proxy recall tasks. Specifically, it generates pseudo-queries that match the granularity of the actual query while pointing to the user's positively interacted items as verifiable proxy targets, evaluating which modality yields better recall performance in analogous scenarios and optimizing the routing strategy in an agentic manner. It then performs routed recall with optimized strategy, enriches results with collaborative items, and ranks candidates by their relevance to both the query and user preferences. Experiments demonstrate that \ourmethod delivers strong performance against state-of-the-art baselines, highlighting the effectiveness and broader potential of adaptive control over modality reliance in multimodal recommendation.
\end{abstract}

\section{Introduction} 
\label{sec:intro} 
\begin{figure*}[t]
    \centering
    \includegraphics[width=0.85\linewidth]{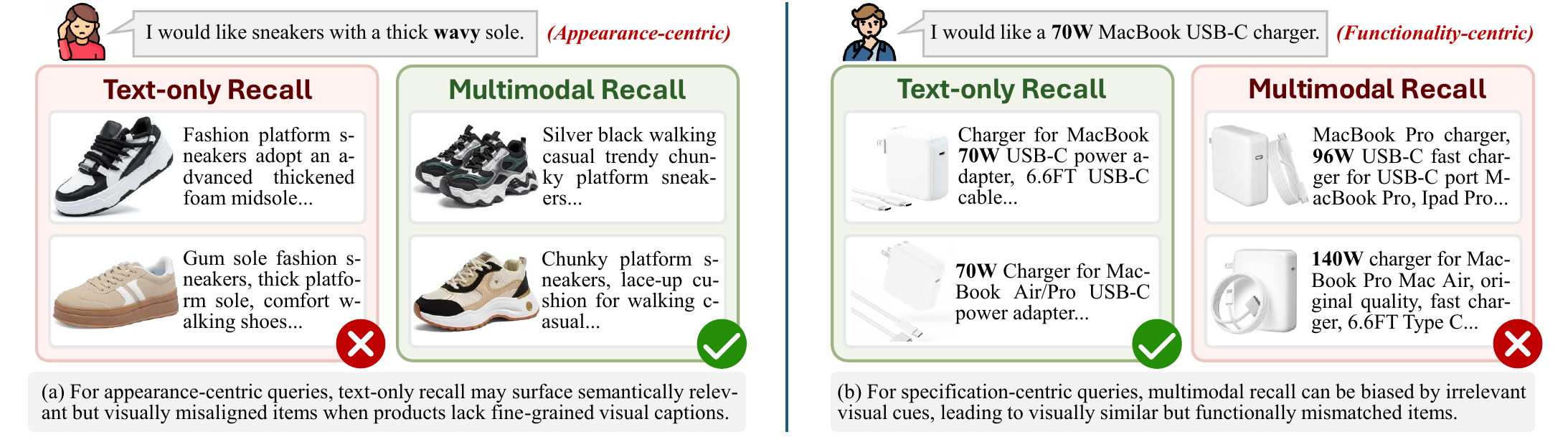}
    \caption{\textbf{The effectiveness of visual information is not universal.} For example: (a) for appearance-centric queries, visuals could capture details that are missing from text; (b) for functionality-centric queries, irrelevant visual cues may introduce noisy features and lead to suboptimal results.}
    \label{fig:moti}
\vspace{-0.22in}
\end{figure*}

In e-commerce, user decisions are naturally shaped by multimodal product information, ranging from textual attributes such as descriptions and specifications to visual characteristics related to aesthetics and appearance~\cite{wang2026mllmrec,ye2025harnessing,fu2025vistawise}. This motivates increasing research efforts in multimodal recommender systems, which have enhanced item profiling and user modeling by incorporating textual and visual information through representation learning~\cite{tao2020mgat,zhai2025simple}, multimodal tokenization~\cite{zhaimultimodal}, and multimodal large language models (MLLMs)~\cite{pomo2025recommender,liu2025onerec,zhang2025improving}. However, the effectiveness of visual information is not universal~\cite{zhu2024bringing,fu2025brainvis}; rather, it depends on the extent to which a product’s visual attributes align with the user’s intent. As illustrated in \autoref{fig:moti}, the visual information of products could be essential for appearance-driven queries such as ``sneakers with a thick wavy sole,'' but may provide little evidence for functionality-driven queries like ``70W MacBook USB-C charger.'' In such cases, product images may even introduce noise by retrieving visually similar yet functionally mismatched items~\cite{zhu2025query,zhu2024bringing}.\footnotetext{Codes are available at \url{https://github.com/RomGai/AdaM-Rec}}.

This query-level variability in modality reliance poses a fundamental challenge for existing multimodal recommenders. Although prior studies have shown the benefits of incorporating both textual and visual information~\cite{lei2026unirec,zhang2024beyond}, they typically treat modality integration as static: textual and visual signals are fused through fixed architectures and globally learned embeddings~\cite{tao2020mgat,zhai2025simple,wei2019mmgcn}, or through predefined retrieval pipelines~\cite{ye2025harnessing,dang2025mllmrec,wu2025refineshot,mei2025a1}. Such designs implicitly assume that the reliance on different modalities remains stable across recommendation scenarios, whereas, as the examples above illustrate, it can vary substantially across queries~\cite{zhu2025query,chen2025haif}. Consequently, fixed modality-integration strategies may yield suboptimal performance in diverse recommendation contexts. Therefore, the key challenge in multimodal recommendation is not only to fuse textual and visual information, but also to dynamically determine the appropriate reliance on each modality for a given query.

To address this challenge, we propose \textbf{\ourmethod}, an LLM-based framework for \textbf{Ada}ptive \textbf{M}odality routing in multimodal \textbf{Rec}ommendation. Instead of assuming that visual signals are always beneficial, \ourmethod adaptively calibrates the reliance on textual and multimodal evidence for user-specific queries. Its core insight is to estimate modality reliability using proxy recall tasks built from the user's historical interactions. Specifically, since the real target of an actual query is unknown, \ourmethod treats the user's positively interacted items as verifiable proxy targets that capture recurring user preferences. It then generates pseudo-queries that match the granularity of the actual query while pointing to these known proxy targets. By comparing the effectiveness of text-based and multimodal recall on proxy targets, \ourmethod iteratively estimates which modality better captures the user’s interests and the actual query intent, thereby optimizing the routing strategy in an agentic manner.

This routing mechanism is built on structured representations of items, users, and their collaborative relations. Specifically, \ourmethod structures multimodal item information and user interaction histories into natural-language item profiles and user preferences, and models collaborative relations via similarities among user preferences. Such representations enable \ourmethod to agentically optimize the modality routing strategy through the native reasoning capabilities of its LLM policy. It then performs routed recall with the optimized strategy, enriches the results with collaborative items, and ranks the final candidates by their relevance to both the query and user preferences. Collectively, \ourmethod achieves better personalized recommendations through user-aware and query-specific routing of modality reliance. Our contributions are summarized as follows:

\begin{itemize}[leftmargin=1.2em, itemsep=0pt, parsep=0pt, topsep=0pt, partopsep=0pt]
    \item We propose \ourmethod, to the best of our knowledge, the first framework for adaptive modality routing in multimodal recommendation. By formulating candidate recall as an agentic self-optimization process, it dynamically calibrates the reliance on textual and multimodal evidence for user-specific queries, leading to improved downstream recommendation performance.
    \item We introduce a novel pseudo-query mechanism for adaptive modality routing. For an actual query, \ourmethod generates granularity-matched pseudo-queries over user historical interactions, constructing proxy recall tasks with verifiable historical targets. This allows it to agentically determine the optimal reliance on different modalities before executing the actual query.
    \item Extensive experiments demonstrate that \ourmethod achieves strong performance compared to state-of-the-art baselines, illustrating the effectiveness and future promise of dynamic routing of modality reliance in multimodal recommender systems.
\end{itemize}

\section{Related Works} 
\label{sec:rw}

\subsection{LLM-based and Agentic Recommendation}

In recent years, LLMs have garnered increasing attention in recommender systems due to their strong contextual modeling and reasoning capabilities~\cite{jiang2025recgpt,zhu2024collaborative,zheng2025universal}. Existing studies have integrated LLMs into classic recommendation pipelines~\cite{shenglanguage,zhou2025openonerec,zhang2022gaze}, either as backbones for sequential recommendation via next-ID prediction~\cite{liu2025onerec,zhu2024cost,liu2025bridging,shi2025llada,wu2026camreasoner}, or as auxiliary modules~\cite{ren2024representation,fu2024dp} to improve collaborative filtering through representation alignment~\cite{hou2023learning,peng2025denoising,ren2025coherence,fu2023sgcn} and knowledge/semantic enhancement~\cite{wang2024llmrg,sun2025llmser,mei2025surveycontextengineeringlarge}. However, recent studies~\cite{zhang2026unleashing,hong2025llm,wang2025enhancing,ge2025focusingcontrastiveattentionenhancing} show that item/semantic IDs and collaborative filtering representations in traditional recommender pipelines are not naturally aligned with the language space of LLMs. As a result, recommender systems may not fully exploit LLMs’ language priors for complex reasoning~\cite{hong2025eager,ye2026align3gr,yuyao2022vision}. This has motivated LLM-based agents~\cite{yang2025agentdr,zhang2024agentcf,xia2026multi,wu2025personalized} that formulate recommendation as an adaptive natural-language decision-making process~\cite{fang2024multi,portugal2024agentic,kong2025think,chen2026shopx}, enabling fine-grained preference modeling through multi-step reasoning and collaboration~\cite{shu2024rah,yu2025intelligent,wang2025tunable,xiao2025mmagentrec}. Inspired by the capacity of agents for context-aware decision-making and iterative optimization~\cite{mei2026gateddifferentiableworkingmemory,fu2026videostir,chen2025tokensnodessemanticguidedmotion}, we cast multimodal integration as an agentic modality-routing problem, enabling dynamic calibration of recall strategies across modalities for each user-specific query.

\subsection{Multimodal Recommendation}

To capture richer item features and user preferences, prior work has incorporated multimodal signals, such as images and text, into recommendation models via contrastive learning~\cite{zhai2025simple,ge2025framemind,fu2025sdr}, multimodal tokenization~\cite{zhaimultimodal}, or adaptive feature fusion~\cite{hu2023adaptive,zhang2026test}. These methods enhance user and item representations beyond purely textual or ID-based signals, demonstrating the value of multimodal information for downstream recommendation tasks~\cite{ye2025harnessing,lei2026unirec,zhang2024defending}. More recently, MLLMs have shown promise in recommendation~\cite{wang2026mllmrec,huang2026dmesr,dang2025mllmrec,ge2026should}, owing to their enhanced visual understanding and inherited reasoning abilities from LLMs. It makes them a natural fit for multimodal item profiling and user  modeling~\cite{wang2025leveraging,wang2025mmsrarec,wang2025multimodal,zhang2025tokenswap}. For example, MLLM-MSR~\cite{ye2025harnessing} explores a multimodal sequential recommendation framework that leverages MLLMs to integrate product's visual and textual information and model user preferences, whereas MLLMRec~\cite{dang2025mllmrec} introduces MLLMs into collaborative modeling to improve multimodal recommendation performance. However, as noted earlier, existing approaches still lack explicit mechanisms to adaptively calibrate the use of multimodal information across different recommendation requests. In this paper, we propose an agentic framework with an adaptive modality routing mechanism that enables the recommender to dynamically regulate the importance of different modalities across recommendation contexts.

\section{\ourmethod}
\label{sec:method}

\begin{figure*}[t]
    \centering
    \includegraphics[width=\linewidth]{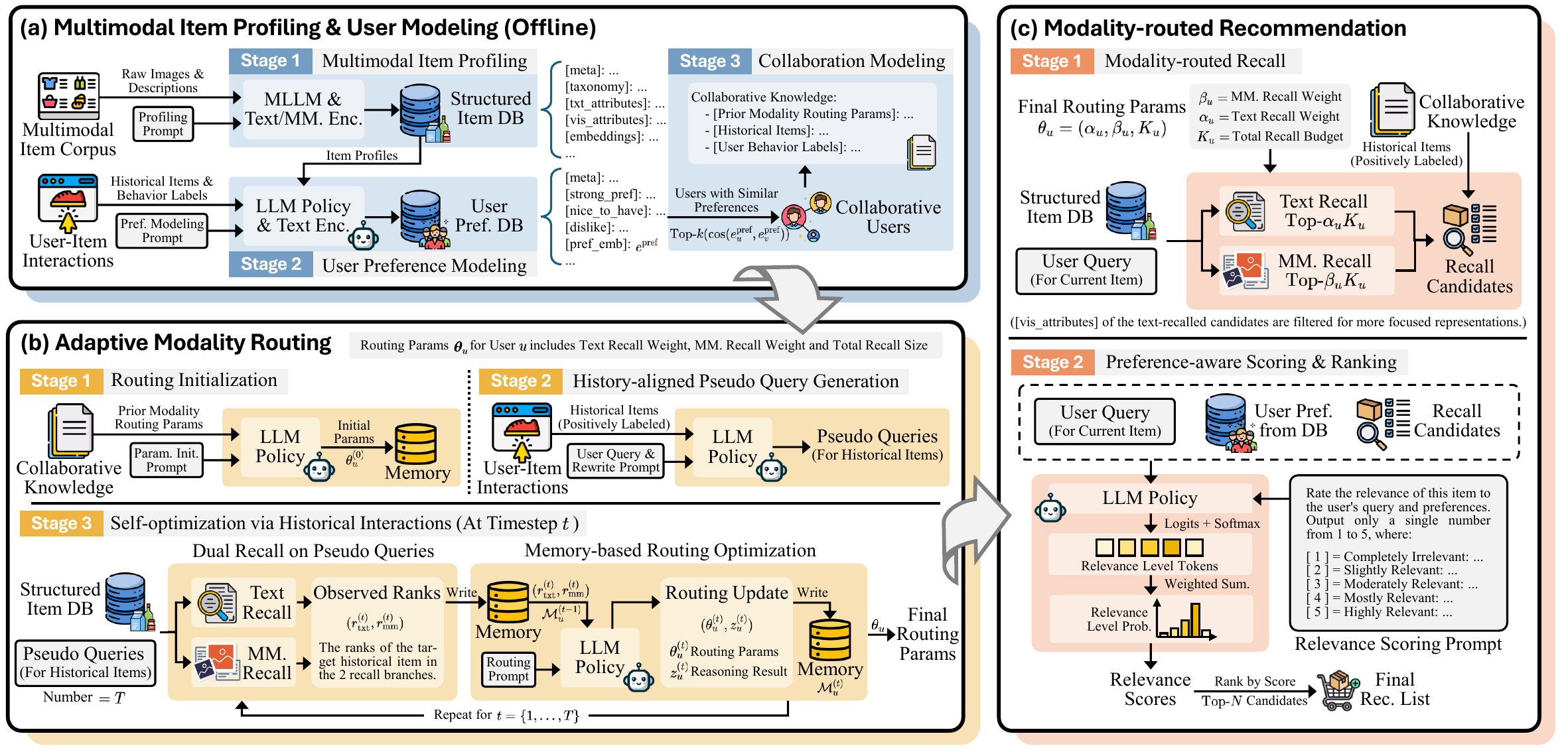}
    \caption{\textbf{Overview of \ourmethod.} The framework consists of three components: (a) multimodal item profiling \& user modeling module, which converts raw images, textual descriptions, and user interaction histories into structured item/preference profiles and collaborative knowledge; (b) adaptive modality routing module, which operates in an agentic manner. It initializes routing from collaborative knowledge and iteratively refines the allocation of text/multimodal recall budget via history-aligned pseudo-queries and memory-based self-optimization; (c) modality-routed recommendation module, which performs text/multimodal recall under the optimized routing strategy, augments candidate items with collaborative knowledge, and ranks them by their relevance to the user’s query and preferences to generate the final recommendation list.}
    \label{fig:fw}
\vspace{-0.22in}
\end{figure*}

\ourmethod is a multimodal recommendation framework equipped with an \textit{adaptive modality routing mechanism}. Given a user-specific query, instead of assuming that visual information is always beneficial for downstream recommendation, \ourmethod dynamically determines the relative reliance on text-based and multimodal recall tools. As illustrated in \autoref{fig:fw}, the framework consists of three modules: (1) \textbf{Multimodal Item Profiling and User Modeling} (Offline), which constructs structured representations for multimodal item profiles, user preferences, and collaborative knowledge; (2) \textbf{Adaptive Modality Routing}, which agentically optimizes the allocation of the recall budget between textual and multimodal branches based on the user’s query, historical interactions, and collaborative knowledge from users with similar preferences; and (3) \textbf{Modality-routed Recommendation}, which recalls candidates under the optimized modality routing schedule, augments them with collaborative candidates, and scores them according to user-specific query and preference relevance to rank the final recommendation list. Collectively, these modules enable an agentic multimodal recommendation framework that adaptively calibrates the reliance on text-based and multimodal recall according to user-specific query, historical interactions, and collaborative knowledge, thereby improving the quality of downstream personalized recommendation. 

\subsection{Multimodal Item Profiling and User Modeling}
\label{sec:profiling}
This module aims to construct structured profiles of multimodal item attributes, infers user preferences from historical interactions, and organizes collaborative knowledge from users with similar preference, thereby providing a foundation for downstream modality routing and personalized recommendation.

\noindent\textbf{Multimodal Profiling for Items.} For each item $i=({v}_i,{l}_i)$ in raw item corpus $\mathcal{I}$, \ourmethod calls an MLLM $\phi_{\text{M}}$ with a profiling prompt $\mathcal{P}_{\text{profile}}$ to extract a structured textual item profile ${s}_i=\phi_{\text{M}}(i,\mathcal{P}_{\text{profile}})$, which summarizes the multimodal attributes from its textual descriptions $\text{l}_i$ and main image $\text{v}_i$. It further computes a text embedding ${e}_i^{\text{txt}}=E_{\text{txt}}({l}_i)$ and a multimodal embedding ${e}_i^{\text{mm}}=E_{\text{mm}}({v}_i,{l}_i)$ for each item via text and multimodal encoders, $E_{\text{txt}}$ and $E_{\text{mm}}$, respectively. These representations form the multimodal item database:
\begin{equation}
\mathcal{D}^{\text{item}}=\{({s}_i,{e}_i^{\text{txt}},{e}_i^{\text{mm}})\}_{i\in\mathcal{I}}.
\end{equation}

\noindent\textbf{Modeling User Preferences from Historical Interactions.} Given a user's historical interaction sequence $u=\{(i_u^t,b_u^t)\}_{t=1}^{T_u}\in\mathcal{U}$ with a length of $T_u$, where $b_u^t\in\{\texttt{pos},\texttt{neg}\}$ denotes the behavior label at step $t$, each interacted item $i_u^t$ is mapped to its corresponding item profile ${s}_{u}^{t}$ in $\mathcal{D}^{\text{item}}$ and construct the user historical behavior profile ${h}_u=\{({s}_{u}^{t}, b_u^t, t)\}_{t=1}^{T_u}$. \ourmethod then leverages its LLM policy $\phi_{\text{L}}$ to infer a structured textual preference profile ${p}_u=\phi_{\text{L}}({h}_u,\mathcal{P}_\text{pref})$ with a preference modeling prompt $\mathcal{P}_\text{pref}$. The resulting profile summarizes recurring preference patterns across multimodal attributes such as style, color, material, silhouette, function, brand, compatibility and other item-level signals. Specifically, we categorize user preferences into strong positive requirements (\texttt{strong\_pref}), weaker preference cues (\texttt{nice\_to\_have}), explicit negative constraints (\texttt{dislike}). After inferring ${p}_u$ from ${h}_u$, \ourmethod writes them into the global user preference profile database:
\begin{equation}
\mathcal{D}^\text{pref}=\{{h}_u,{p}_u,{e}^\text{pref}_u\}_{u\in\mathcal{U}},
\end{equation}
where ${e}^{\text{pref}}_u = E_{\text{txt}}({p}_u)$ is the textual embedding of the user's preference profile, thereby incorporating the current user into a retrievable collaborative space for other users.

\noindent\textbf{Preference-based Collaboration Modeling.} To model collaborative relations among users, \ourmethod first computes the cosine similarity between the current user’s preference embedding ${e}^\text{pref}_u$ and those of others in $\mathcal{D}^\text{pref}$. It then retrieves the top-$K_c$ most similar users as the collaborative user set $\mathcal{N}_u$. For each collaborative user $v\in\mathcal{N}_u$, the collaborative knowledge consists of $({\theta}_v,{s}_v,{b}_v)$. Specifically, ${\theta}_v$ denotes the previously optimized modality routing parameters of user $v$ (Sec.~\ref{sec:modality_tuning}), if available. Meanwhile, ${s}_v$ and ${b}_v$ denote the historically interacted items of user $v$ and the corresponding behavior labels, respectively. The collaborative knowledge for the current user can be represented as:
\begin{equation}
\mathcal{C}_u=\{({\theta}_v,{s}_v,{b}_v)\}_{v\in\mathcal{N}_u}.
\end{equation}

\subsection{Adaptive Modality Routing}
\label{sec:modality_tuning}
This module optimizes the modality-routing strategy for user-specific recommendation requests by considering the current actual query, historical interactions, and collaborative knowledge. It introduces a novel routing mechanism that agentically constructs proxy recall tasks through pseudo-queries aligned with the user’s historically interacted targets and the granularity of the actual query. By evaluating text-based and multimodal recall branches on these proxy tasks, it adaptively allocates modality weights and recall budget at inference time for the actual query.

\noindent\textbf{Initialization from Collaborative Priors.} Given the collaborative knowledge $\mathcal{C}_u$ of the current user, the LLM policy $\phi_{\mathrm{L}}$ leverages the parameter-initialization prompt $\mathcal{P}_{\mathrm{pi}}$ to analyze common priors in the modality-routing parameters of collaborative users, thereby estimating the initial routing state ${\theta}^{(0)}_{u}$ for the user:
\begin{equation}
{\theta}^{(0)}_{u}
=
\left(
\alpha^{(\mathrm{0})}_u,
\beta^{(\mathrm{0})}_u,
K^{(\mathrm{0})}_{u}
\right)
=
\phi_{\mathrm{L}}(\mathcal{C}_u, \mathcal{P}_{\mathrm{pi}}), 
\end{equation}
where $\alpha^{(\mathrm{0})}_u$ and $\beta^{(\mathrm{0})}_u$ denote the weights for text-based and multimodal recall branches at timestep $t=0$, respectively, while $K^{(\mathrm{0})}_{u}$ denotes the initial recall budget. The weights satisfy $\alpha^{(0)}_u+\beta^{(0)}_u=1$. Subsequently, this initial state is stored in memory $\mathcal{M}_u^{(0)}=\{{\theta}^{(0)}_{u}\}$ to facilitate subsequent iterative optimization, where $K_{u}$ is constrained to be no greater than a threshold constant $\tau$.


\noindent\textbf{Self-optimization via Historical Interactions.} To make modality routing personalized rather than purely heuristic, \ourmethod optimizes routing parameters based on current user’s own historical interacted items. The key motivation is that, for a user's actual query, the target item is unknown, making it difficult to directly determine the effectiveness of text-based and multimodal recall for the query. In contrast, related and positively interacted historical items are observed choices from the same user and thus provide user-specific, verifiable proxy targets that reflect the user’s recurring preference patterns, facilitating the optimization of routing parameters.

Specifically, given the current actual query $q$, the LLM policy $\phi_{\mathrm{L}}$ uses a rewriting prompt $\mathcal{P}_{\mathrm{rw}}$ to reformulate it into a set of history-aligned pseudo-queries $\widetilde{\mathcal{Q}}_u=\phi_{\mathrm{L}}(q,{h}_u, \mathcal{P}_{\mathrm{rw}})=\{\tilde{q}_1,\tilde{q}_2,\dots,\tilde{q}_{T_f}\}$, where each pseudo-query $\tilde{q}_t$ is expressed at a granularity similar to $q$ but targets one known and related historical item from the user’s interaction history. During this process, $\phi_{\mathrm{L}}$ also filters out historical items that are irrelevant with respect to $q$ or negatively interacted, resulting in a filtered interaction sequence of length $T_f$. In this way, \ourmethod constructs proxy retrieval tasks that preserve the search form of the current recommendation request while replacing unknown target with verifiable user-preferred items. Subsequently, it iteratively optimizes the recall strategy based on these pseudo-queries and history items. At timestep $t>0$, it invokes the text-based and multimodal recall on $\tilde{q}_t$, and obtains the rank positions of the target item $i_u^t$ in the corresponding retrieval results, denoted by $r_{\text{txt}}^{(t)}$ and $r_{\text{mm}}^{(t)}$, respectively. These relative ranks provide evidence about which modality better matches the user’s preference structure under analogous search conditions. Based on these observations, the previous status stored in the memory $\mathcal{M}_u^{(t-1)}$, and the modality-routing prompt $\mathcal{P}_{\mathrm{mr}}$ that guides the reasoning process, the LLM policy $\phi_{\mathrm{L}}$ infers a refined parameter state for the next iteration:
\begin{equation}
\left(
{\theta}_u^{(t)},
{z}_u^{(t)}
\right)
=
\phi_{\mathrm{L}}
\big(
r_{\text{txt}}^{(t)}, r_{\text{mm}}^{(t)},
\mathcal{M}_u^{(t-1)},
\mathcal{P}_{\mathrm{mr}}
\big),
\quad T_f \ge t > 0,
\end{equation}
where ${z}_u^{(t)}$ denotes the policy’s reasoning summary for the update. To preserve the optimization trajectory, the result of each iteration is written into the user-specific memory:
\begin{equation}
\mathcal{M}_u^{(t)}
=
\mathcal{M}_u^{(t-1)}
\cup
\left\{
\left(
r_{\text{txt}}^{(t)},\,
r_{\text{mm}}^{(t)},\,
{\theta}_u^{(t)},\,
{z}_u^{(t)}
\right)
\right\}.
\end{equation}
After processing all pseudo-queries, the final modality routing parameters for the current user and query are taken as ${\theta}_u={\theta}_u^{(T)}$, which are then used to control downstream recall for the real query. Moreover, the routing parameter ${\theta}_u$ is written back into the current user’s preference profile $\mathcal{D}^{\mathrm{pref}}[u] \leftarrow \{{\theta}_u\}$, enabling its reuse as collaborative knowledge for other users. 

\subsection{Modality-Routed Recommendation}
\label{sec:recall_rerank}
This module performs downstream retrieval using the final modality routing parameters optimized in the previous stage. It allocates recall budget across text-based and multimodal branches, augments the recalled candidates with collaborative evidence from preference-similar users, and applies preference-aware scoring to rank a recommendation list aligned with both query intent and the user’s preferences.

\noindent\textbf{Modality-Routed Recall.} Given the query $q$ and the item database $\mathcal{D}^{\text{item}}=\{({s}_i,{e}_i^{\text{txt}},{e}_i^{\text{mm}})\}_{i\in\mathcal{I}}$, \ourmethod invokes downstream text-based and multimodal recall with the optimized final routing parameters ${\theta}_u=(\alpha_u,\beta_u,K_u)$ to construct an initial recall set $\mathcal{R}^{\text{init}}_{u,q}$, by taking the Top-$\alpha_u K_u$ items whose ${e}_i^{\text{txt}}$ are most similar to query's embedding $E_{\text{txt}}(q)$ in textual space, and the Top-$\beta_u K_u$ items whose ${e}_i^{\text{mm}}$ are most similar to query's embedding $E_{\text{mm}}(q)$ in multimodal space. For candidates recalled by text-based retrieval, visual attributes in their profiles are selectively filtered to produce cleaner and more focused candidate representations.

\noindent\textbf{Collaborative Recall Augmentation.}
To complement the recall results, we further incorporate positive historical items from preference-similar users as collaborative candidates:
\begin{equation}
\mathcal{R}^{\mathrm{co}}_{u}
=
\left\{
{s}_v^t
\;\middle|\;
v\in\mathcal{N}_u,\; 1\le t\le T_v,\; b_v^t=\texttt{pos}
\right\},
\qquad
\mathcal{R}^{\mathrm{aug}}_{u}
=
\mathcal{R}^{\mathrm{init}}_{u}\cup \mathcal{R}^{\mathrm{co}}_{u}.
\end{equation}
This augmentation complements user-specific modality routing by injecting collaborative evidence from preference-similar users, thereby improving recall coverage and increasing the likelihood of retrieving items that align with the current user’s latent preferences.

\noindent\textbf{Preference-aware Scoring \& Ranking.} 
Given the augmented candidate set, \ourmethod estimates a preference-aware relevance score for each candidate. Conditioned on the query $q$, the user preference profile ${p}_u$, the candidate profile ${s}_i$, and the scoring prompt $\mathcal{P}_{\mathrm{sc}}$, the LLM policy $\phi_{\mathrm{L}}$ outputs a softmax-normalized probability distribution over discrete relevance-level tokens $\ell \in \{1,2,3,4,5\}$:
\begin{equation}
P_{\mathrm{R}}\!\left(\ell \mid q,p_u,{s}_i,\mathcal{P}_{\mathrm{sc}}\right)
=
\frac{
\exp\!\left(
\phi_{\mathrm{L}}(\ell \mid q,p_u,{s}_i,\mathcal{P}_{\mathrm{sc}})
\right)
}{
\sum_{k=1}^{5}
\exp\!\left(
\phi_{\mathrm{L}}(k \mid q,p_u,{s}_i,\mathcal{P}_{\mathrm{sc}})
\right)
},
\end{equation}
where $\phi_{\mathrm{L}}(\cdot)$ denotes the logits produced by the LLM policy, and $\mathcal{P}_{\mathrm{sc}}$ guides the policy to assess whether each candidate satisfies the query intent $q$ while aligning with the user preferences in $p_u$. The final candidate score is computed as the expected relevance level:
\begin{equation}
Score_i
=
\sum_{\ell=1}^{5}
\ell \cdot
P_{\mathrm{R}}\!\left(\ell \mid q,p_u,s_i,\mathcal{P}_{\mathrm{sc}}\right). 
\end{equation}
This formulation assigns each candidate a continuous score based on query intent, user preferences, and item profiles. Candidates with higher scores are deemed more relevant and better aligned with user's queries and personalized preferences, with the top-$N$ forming the final recommendation list.

\section{Experiments}
\label{sec:exp}

\subsection{Dataset and Implementation}
\noindent\textbf{Datasets.} To ensure fair comparisons with prior work, we strictly follow the experimental settings of the corresponding baselines in each recommendation task. For query-based recommendation, we use the user-query datasets constructed by TAIRA~\cite{yu2026thought} on three Amazon Review subsets~\cite{mcauley2015image,he2016ups}: Amazon Clothing \& Shoes, Amazon Beauty, and Amazon Music. These datasets contain 94,328, 91,263, and 302,803 items, respectively, with 361, 374, and 393 carefully curated user-query pairs. For the extended preference-based recommendation, we adopt the evaluation protocol of MLLM-MSR~\cite{ye2025harnessing} on Amazon Video Games and Amazon Baby Products. After standard 5-core preprocessing~\cite{ye2025harnessing}, the two datasets include 12,559 and 19,945 users, and 18,017 and 24,847 items, respectively.

\noindent\textbf{Implementation.} \ourmethod uses Qwen3-8B~\cite{yang2025qwen3} as the default LLM policy. It employs Qwen3-Emb.-0.6B~\cite{zhang2025qwen3} as the text encoder $E_\text{txt}$, Qwen3-VL-Emb.-2B~\cite{li2026qwen3} as the multimodal encoder $E_\text{mm}$, and Qwen3-VL-8B-Instruct~\cite{bai2025qwen3} as the MLLM $\phi_\text{M}$ for multimodal profiling. For a fair comparison, we upgrade the backbones of earlier baselines that adopt weaker (M)LLMs or encoders to the same setting as ours. All open-source models are deployed on A100 80GB GPUs. Most hyperparameters are automatically optimized by the policy, except for the collaborative users number $K_c=5$ and maximum recall budget threshold $\tau=1000$, following the analysis in Sec.\ref{sec:dis}. Unless otherwise stated, we report the median over five trials, perform ablations on Amazon Beauty, and randomly sample 20\% of users as the test set for preference-based recommendation. Prompts involved in Sec.\ref{sec:method} and profile examples are provided in Appendices~\ref{sec:prompt} and \ref{sec:formats}, respectively.

\subsection{Comparison Against Other Methods}
\noindent\textbf{Query-based Recommendation.}
As shown in \autoref{tab:comp1}, we evaluate \ourmethod on query-based personalized recommendation task, where the system recalls and ranks items from the full corpus based on the user's query and preferences inferred from historical interactions. Following TAIRA~\cite{yu2026thought}, we assess full-corpus groundtruth item recommendation using Hit Rate (HR) and Normalized Discounted Cumulative Gain (NDCG). We compare against the same categories of baselines as TAIRA, including classical retrieval/reranking models~\cite{trotman2014improvements,li2023making,chen2024m3}, general agentic frameworks~\cite{yao2022react,shinn2023reflexion} instantiated under the TAIRA setting, and state-of-the-art (SOTA) interactive agentic recommender~\cite{huang2025recommender,fang2024multi,wang2024macrec,yu2026thought,xia2026multi} compatible with this task. Results are taken from open-source codebases or reproduced using official settings. \ourmethod consistently outperforms prior baselines on query-based recommendation tasks, achieving average gains of 14.1\% in HR and 15.9\% in NDCG, demonstrating its effectiveness.
\begin{table*}[t]
\centering
\small
\setlength{\tabcolsep}{5pt}
\caption{{\textbf{Comparison with query-based recommendation baselines.} ``@K'' denotes the HR and NDCG at top-K. Bold and underlined values indicate the best and second-best results, respectively.}}
\resizebox{\linewidth}{!}{
\begin{tabular}{l cccc cccc cccc}
\toprule
\multirow{3}{*}{\textbf{Method}} &
\multicolumn{4}{c}{\textbf{Amazon Beauty}} &
\multicolumn{4}{c}{\textbf{Amazon Clothing}} &
\multicolumn{4}{c}{\textbf{Amazon Music}} \\
[-1pt] \cmidrule(lr){2-5} \cmidrule(lr){6-9} \cmidrule(lr){10-13} \noalign{\vskip -1pt}

& \multicolumn{2}{c}{\textbf{HR $\uparrow$}} &
\multicolumn{2}{c}{\textbf{NDCG $\uparrow$}} &
\multicolumn{2}{c}{\textbf{HR $\uparrow$}} &
\multicolumn{2}{c}{\textbf{NDCG $\uparrow$}} &
\multicolumn{2}{c}{\textbf{HR $\uparrow$}} &
\multicolumn{2}{c}{\textbf{NDCG $\uparrow$}} \\ [-1pt] 
\cmidrule(lr){2-3} \cmidrule(lr){4-5}
\cmidrule(lr){6-7} \cmidrule(lr){8-9}
\cmidrule(lr){10-11} \cmidrule(lr){12-13} \noalign{\vskip -1pt}

& \textbf{@20} & \textbf{@40}
& \textbf{@20} & \textbf{@40}
& \textbf{@20} & \textbf{@40}
& \textbf{@20} & \textbf{@40}
& \textbf{@20} & \textbf{@40}
& \textbf{@20} & \textbf{@40} \\ \noalign{\vskip -1pt}
\midrule
BM25~\cite{trotman2014improvements}  
& 0.0348 & 0.0508 & 0.0166 & 0.0199
& 0.0194 & 0.0277 & 0.0080 & 0.0096
& 0.0382 & 0.0483 & 0.0183 & 0.0203 \\

BGE-M3~\cite{li2023making}
& 0.0374 & 0.0561 & 0.0173 & 0.0211
& 0.0249 & 0.0388 & 0.0093 & 0.0122
& 0.0458 & 0.0611 & 0.0239 & 0.0269 \\

BGE-Reranker~\cite{chen2024m3}
& 0.0401 & 0.0615 & 0.0184 & 0.0227
& 0.0277 & 0.0443 & 0.0137 & 0.0171
& 0.0483 & 0.0687 & 0.0204 & 0.0245 \\

\midrule

ReAct~\cite{yao2022react}
& 0.0909 & 0.1096 & 0.0369 & 0.0408
& 0.0499 & 0.0720 & 0.0278 & 0.0322
& 0.0789 & 0.0967 & 0.0411 & 0.0448 \\

Reflexion~\cite{shinn2023reflexion}
& 0.0963 & 0.1230 & 0.0491 & 0.0547
& 0.0582 & 0.0776 & 0.0308 & 0.0349
& 0.0840 & 0.0992 & 0.0494 & 0.0525 \\

\midrule

InteRecAgent~\cite{huang2025recommender}  
& 0.0989 & 0.1257 & 0.0454 & 0.0508
& 0.0360 & 0.0693 & 0.0141 & 0.0207
& 0.0814 & 0.0916 & 0.0632 & 0.0651 \\

MACRS~\cite{fang2024multi} 
& 0.1016 & 0.1310 & 0.0473 & 0.0533
& 0.0443 & 0.0554 & 0.0175 & 0.0197
& 0.0585 & 0.0687 & 0.0221 & 0.0243 \\

MACRec~\cite{wang2024macrec}
& 0.1471 & 0.1765 & \underline{0.0803} & 0.0865
& 0.0554 & 0.0665 & 0.0283 & 0.0305
& 0.0763 & 0.0891 & 0.0569 & 0.0595 \\

MACF~\cite{xia2026multi} 
& \underline{0.1551} & 0.1872 & 0.0725 & 0.0791
& 0.0886 & 0.1191 & \underline{0.0510} & \underline{0.0574}
& 0.1221 & 0.1374 & 0.0679 & 0.0710 \\

TAIRA~\cite{yu2026thought} 
& 0.1497 & \underline{0.1925} & 0.0798 & \underline{0.0886}
& \underline{0.0914} & \underline{0.1274} & 0.0457 & 0.0532
& \underline{0.1349} & \underline{0.1450} & \underline{0.0939} & \underline{0.0960} \\

\specialrule{\lightrulewidth}{1.5pt}{0pt}
\rowcolor[rgb]{0.80,0.90,0.95} \rule[-0.9ex]{0pt}{3.3ex} \textbf{AdaM-Rec} 
& \textbf{0.1711} & \textbf{0.2193} & \textbf{0.0979} & \textbf{0.1077}
& \textbf{0.1108} & \textbf{0.1468} & \textbf{0.0574} & \textbf{0.0648}
& \textbf{0.1501} & \textbf{0.1629} & \textbf{0.1063} & \textbf{0.1089} \\
\tightbottomrule
\end{tabular}
}
\label{tab:comp1}
\vspace{-0.18in}
\end{table*}


\begin{table*}[t]
\centering
\small
\setlength{\tabcolsep}{11pt}
\caption{{\textbf{Comparison with preference-based recommendation baselines.} ``@K'' denotes HR and NDCG at top-K. Bold and underlined values indicate best and second-best results.}}
\resizebox{\linewidth}{!}{
\begin{tabular}{l cccc cccc}
\toprule
\multirow{3}{*}{\textbf{Method}} &
\multicolumn{4}{c}{\textbf{Amazon Video Games}} &
\multicolumn{4}{c}{\textbf{Amazon Baby Products}} \\
[-1pt] \cmidrule(lr){2-5} \cmidrule(lr){6-9} \noalign{\vskip -1pt}

& \multicolumn{2}{c}{\textbf{HR $\uparrow$}} &
\multicolumn{2}{c}{\textbf{NDCG $\uparrow$}} &
\multicolumn{2}{c}{\textbf{HR $\uparrow$}} &
\multicolumn{2}{c}{\textbf{NDCG $\uparrow$}} \\ [-1pt] 
\cmidrule(lr){2-3} \cmidrule(lr){4-5}
\cmidrule(lr){6-7} \cmidrule(lr){8-9} \noalign{\vskip -1pt}

& \textbf{@20} & \textbf{@40} & \textbf{@20} & \textbf{@40}
& \textbf{@20} & \textbf{@40} & \textbf{@20} & \textbf{@40} \\ \noalign{\vskip -1pt}
\midrule
GRU4Rec~\cite{hidasi2015session}
& 0.0587 & 0.0865 & 0.0326 & 0.0383
& 0.0391 & 0.0704 & 0.0218 & 0.0282 \\

SASRec~\cite{kang2018self}
& 0.0655 & 0.0976 & 0.0343 & 0.0410
& 0.0424 & 0.0797 & 0.0236 & 0.0312 \\

MGAT~\cite{tao2020mgat}
& 0.0611 & 0.0921 & 0.0330&0.0394 
&0.0416  &0.0762  &0.0228  &0.0299  \\

MMGCN~\cite{wei2019mmgcn}
& 0.0647 & 0.1012 & 0.0347 & 0.0422
& 0.0429 & 0.0822 & 0.0241 & 0.0322 \\

MMSR~\cite{hu2023adaptive}
& 0.0694 & 0.1067 &0.0359  & 0.0436
& 0.0547 & 0.0905 & 0.0256 & 0.0329 \\

Trans2D~\cite{singer2022sequential}
& 0.0536 & 0.0798 & 0.0283 & 0.0337
& 0.0346 & 0.0569 & 0.0178 & 0.0224 \\

\midrule

MLLM-MSR~\cite{ye2025harnessing}
& 0.0706 & 0.1099 & 0.0367 & 0.0478
& 0.0448 & 0.0787 & 0.0208 & 0.0277 \\

AB-Rec~\cite{wu2025aligning}
& 0.0734 & 0.1179 & 0.0296 & 0.0387
& 0.0652 & 0.1073 & 0.0251 & 0.0337 \\

LLaRa~\cite{liao2024llara}
& 0.0786 & 0.1083 & 0.0329 & 0.0396
& 0.0531 & 0.0895 & 0.0201 & 0.0276 \\

LLMRec~\cite{wei2024llmrec}
& 0.0849 & 0.1187 & 0.0401 & 0.0473
& 0.0622 & 0.0950 & 0.0243 & 0.0315 \\

TallRec~\cite{bao2023tallrec}
& 0.0758 & 0.1052 & 0.0334 & 0.0394
& 0.0479 & 0.0862 & 0.0191 & 0.0270 \\

SPRec~\cite{gao2025sprec}
&  0.0853& 0.1159 & 0.0390 & 0.0452
& 0.0454 & 0.0822 &0.0204  &0.0279  \\

RecZero~\cite{kong2025think}
& 0.0873 & 0.1143 & 0.0362 & 0.0421
& 0.0509 & 0.0920 & 0.0196 & 0.0286 \\

MLLMRec-R1~\cite{wang2026mllmrec}
& \underline{0.0940} & \underline{0.1226} & \textbf{0.0434} & \underline{0.0481}
& \underline{0.0679} & \underline{0.1083} & \underline{0.0261} & \underline{0.0344} \\

\specialrule{\lightrulewidth}{1.5pt}{0pt}
\rowcolor[rgb]{0.80,0.90,0.95} \rule[-0.9ex]{0pt}{3.3ex} \textbf{AdaM-Rec}
& \textbf{0.0967} & \textbf{0.1267} & \underline{0.0422} & \textbf{0.0484}
& \textbf{0.0702} & \textbf{0.1138} & \textbf{0.0264} & \textbf{0.0351} \\
\tightbottomrule
\end{tabular}
}
\label{tab:comp2}
\vspace{-0.22in}
\end{table*}

\noindent\textbf{Preference-based Recommendation.}
As shown in \autoref{tab:comp2}, we further evaluate \ourmethod on the query-free preference-based sequential recommendation task, which predicts the next interacted item based solely on user preferences inferred from historical interactions. To adapt \ourmethod to this setting, we replace the query with the user's current preference for modality routing and downstream retrieval. For each historical item involved in modality routing, its corresponding preference is constructed using only preceding visible interactions, while strictly preserving the format and granularity of the target preference. Following the more challenging MLLM-MSR protocol, we report HR and NDCG by ranking one ground truth item against 1,000 randomly sampled items for each test instance. We compare with baselines aligned with MLLM-MSR~\cite{ye2025harnessing}, including classical and multimodal sequential recommenders~\cite{hidasi2015session,kang2018self,wei2019mmgcn,tao2020mgat,singer2022sequential,hu2023adaptive} and (M)LLM-based methods~\cite{wu2025aligning,liao2024llara,wei2024llmrec,bao2023tallrec,gao2025sprec,kong2025think,ye2025harnessing,wang2026mllmrec}. Results are obtained from open-source codebases. Although this task is not the primary query-based recommendation scenario targeted by \ourmethod, it still achieves competitive performance against SOTA baselines, further demonstrating its generalizability.

\subsection{Ablation Studies}


\noindent\textbf{LLM Policy.} The first group in \autoref{tab:ab1} shows that stronger policy improve modality routing and preference judgment, leading to better recall and ranking performance, though with diminishing gains. Therefore, we adopt Qwen3-8B as the default policy for its cost-performance balance.

\begin{table*}[t]
\centering
\small
\setlength{\tabcolsep}{14pt}
\renewcommand{\arraystretch}{0.96}
\caption{{\textbf{Ablation studies on agent policy and core components.} The metric ``Recall'' denotes the HR of candidates recalled with routed parameters.
``@K'' denotes the HR and NDCG scores computed over top-K ranked items.
Bold highlights the default config and the strongest policy variant.}}
\label{tab:ab1}


\begin{adjustbox}{max width=\linewidth}
\begin{tabular}{l c cccc}
\toprule
\multirow{2}{*}{\textbf{Method}}
& \multirow{2}{*}{\textbf{Recall} $\uparrow$}
& \multicolumn{2}{c}{\textbf{HR $\uparrow$}}
& \multicolumn{2}{c}{\textbf{NDCG $\uparrow$}} \\[-1pt]
\cmidrule(lr){3-4} \cmidrule(lr){5-6}
\noalign{\vskip -1pt}
& & \textbf{@20} & \textbf{@40} & \textbf{@20} & \textbf{@40} \\ [-1pt]

\midrule

\multicolumn{1}{l}{\hspace{-0.35em}\textbf{Full (Qwen3-8B policy)}}
& \textbf{0.4278} & \textbf{0.1711} & \textbf{0.2193} & \textbf{0.0979} & \textbf{0.1077} \\

\specialrule{\lightrulewidth}{1pt}{0pt}
\groupheader{LLM Policy}
Qwen3-32B & \textbf{0.4358} & \textbf{0.1765} & \textbf{0.2246} & \textbf{0.1017} & \textbf{0.1116}\\

\specialrule{\lightrulewidth}{0pt}{0pt}
\groupheader{Adaptive Routing Mechanism}
w/o Adaptive Modality Routing & 0.3743 & 0.1524 & 0.1898 & 0.0787 & 0.0866\\
w/o Collaborative Priors Initialization & 0.4171 & 0.1684 & 0.2139 & 0.0969 & 0.1065\\
w/o Pseudo-query \& Self-optimization & 0.3850 & 0.1578 & 0.1979 & 0.0904 & 0.0988\\
w/o Memory & 0.3930 & 0.1631 & 0.2032 & 0.0938 & 0.1022\\

\specialrule{\lightrulewidth}{0pt}{0pt}
\groupheader{Candidate Recall}
w/o Collaborative Recall Augmentation & 0.4091 & 0.1658 & 0.2086 & 0.0964 & 0.1054\\
w/o Text-based Recall & 0.3155 & 0.1390 & 0.1631 & 0.0726 & 0.0774\\
w/o Multimodal Recall & 0.3422 & 0.1471 & 0.1765 & 0.0768 & 0.0829\\

\specialrule{\lightrulewidth}{0pt}{0pt}
\groupheader{Preference-aware Ranking}
w/o Attribute Filtering & 0.4278 & 0.1551 & 0.2059 & 0.0743 & 0.0848\\
w/o Preference-aware Scoring & 0.4278 & 0.1337 & 0.1684 & 0.0576 & 0.0647\\

\bottomrule
\end{tabular}
\end{adjustbox}

\vspace{-0.2in}
\end{table*}


\begin{table*}[t]
\centering
\small
\setlength{\tabcolsep}{12pt}
\renewcommand{\arraystretch}{0.96}
\caption{{\textbf{Ablations of embedding backbones within different recall branches.} ``Recall'' denotes HR of candidates recalled with routed parameters. ``@K'' denotes the HR and NDCG at top-K.}}
\resizebox{\linewidth}{!}{
\begin{tabular}{ll c cccc}
\toprule
\multirow{2}{*}{\textbf{Text Encoder}} &
\multirow{2}{*}{\textbf{Multimodal Encoder}} &
\multirow{2}{*}{\textbf{Recall} $\uparrow$} &
\multicolumn{2}{c}{\textbf{HR $\uparrow$}} &
\multicolumn{2}{c}{\textbf{NDCG $\uparrow$}} \\ [-1pt] \cmidrule(lr){4-5} \cmidrule(lr){6-7} \noalign{\vskip -1pt}
& & & \textbf{@20} & \textbf{@40} & \textbf{@20} & \textbf{@40} \\  [-1pt]  \noalign{\vskip -1pt}
\midrule

& SigLIP2-L/16 & 0.3289 & 0.1417 & 0.1711 & 0.0745 & 0.0809 \\
SigLIP2-L/16 & Qwen3-VL-Emb.-2B & 0.3904 & 0.1631 & 0.2032 & 0.0865 & 0.0952 \\
& Qwen3-VL-Emb.-8B & 0.4091 & 0.1684 & 0.2112 & 0.0879 & 0.0972 \\
\midrule

& SigLIP2-L/16 & 0.3449 & 0.1471 & 0.1791 & 0.0815 & 0.0884 \\
Qwen3-Emb.-0.6B & Qwen3-VL-Emb.-2B & 0.4278 & 0.1711 & 0.2193 & 0.0979 & 0.1077 \\
& Qwen3-VL-Emb.-8B & 0.4412 & 0.1738 & 0.2246 & 0.0972 & 0.1083 \\
\midrule

& SigLIP2-L/16 & 0.3583 & 0.1497 & 0.1845 & 0.0828 & 0.0904 \\
Qwen3-Emb.-4B & Qwen3-VL-Emb.-2B & 0.4331 & 0.1711 & 0.2219 & 0.0971 & 0.1082 \\
& Qwen3-VL-Emb.-8B & 0.4492 & 0.1764 & 0.2273 & 0.0981 & 0.1091 \\
\bottomrule
\end{tabular}
}
\label{tab:ab2}
\vspace{-0.22in}
\end{table*}

\noindent\textbf{Adaptive Routing Mechanism.} The second group in \autoref{tab:ab1} examines the routing-related components. ``w/o Adaptive Modality Routing'' replaces adaptive routing with the optimal static parameters in  \autoref{fig:dis}, where the text-based/multimodal recall weights are both 0.5, and the total budget is 800, causing a clear performance drop. This highlights the importance of dynamically routing modality reliance across user-specific queries. ``w/o Collaborative Priors Initialization'' removes collaborative-prior-based routing initialization and uses the optimal parameters of static strategy as initial state instead, leading to slight degradation, which verifies its effectiveness as warm start. ``w/o Pseudo-query \& Self-optimization'' keeps only collaborative initialization for each user and shows a larger decline, demonstrating the importance of pseudo-query-based self-optimization for adapting routing parameters to actual query intent. ``w/o Memory'' further degrades performance, confirming that memory helps stabilize and improve self-optimization by reusing useful past routing experiences.

\noindent\textbf{Candidate Recall.} The third group in \autoref{tab:ab1} evaluates the recall-stage design. ``w/o Collaborative Recall Augmentation'' removes historical items from collaborative users, reducing both candidate recall and downstream ranking performance, showing that collaborative knowledge could improve candidate coverage. ``w/o Text-based/Multimodal Recall'' uses only one recall branch with the same total budget as static setting above, which also makes adaptive modality routing invalid. Their substantial degradation indicates the complementarity and necessity of text-based/multimodal recall. The larger drop from removing text-based recall suggests that it may act as the primary semantic anchor, while multimodal recall provides additional cues that cannot be fully captured by text alone.

\noindent\textbf{Preference-aware Ranking.} The last group in \autoref{tab:ab1} analyzes the ranking-stage components. ``w/o Attribute Filtering'' removes visual-attribute filtering from text-recalled candidate profiles, leading to lower downstream HR and NDCG and validating its role in maintaining cleaner, task-relevant representations for improving preference-aware ranking. ``w/o Preference-aware Scoring'' adopts recall-stage embedding similarity for downstream ranking and causes a drop on ranking performance. This confirms that effective recommendation depends not only on retrieving relevant candidates, but also on accurately ranking them according to the current query intent and user preferences.

\noindent\textbf{Embedding backbones.}
As shown in \autoref{tab:ab2}, we conduct ablations on the textual and multimodal embedding backbones used for recall. The results show that stronger encoders generally improve candidate recall, but the gains do not scale proportionally. Scaling up the multimodal encoder brings slightly larger improvements than enlarging the text encoder, suggesting that a compact text encoder is sufficient for text-based recall in our setting, while multimodal recall is more sensitive to encoder capacity. Moreover, using SigLIP2~\cite{tschannen2025siglip} as a vision-only alternative to the multimodal encoder leads to a clear performance drop compared with vision-language encoders such as Qwen3-VL-Embedding. This suggests that the recall benefits from combining visual cues with language-based semantics, such as function, brand, and other fine-grained purchase constraints~\cite{zhu2024bringing}, rather than relying on visual features alone. Overall, we adopt Qwen3-Emb.-0.6B and Qwen3-VL-Emb.-2B as the default configuration for a better efficiency-performance trade-off.

\subsection{Discussion}
\label{sec:dis}

\begin{figure*}[t]
    \centering
    \includegraphics[width=\linewidth]{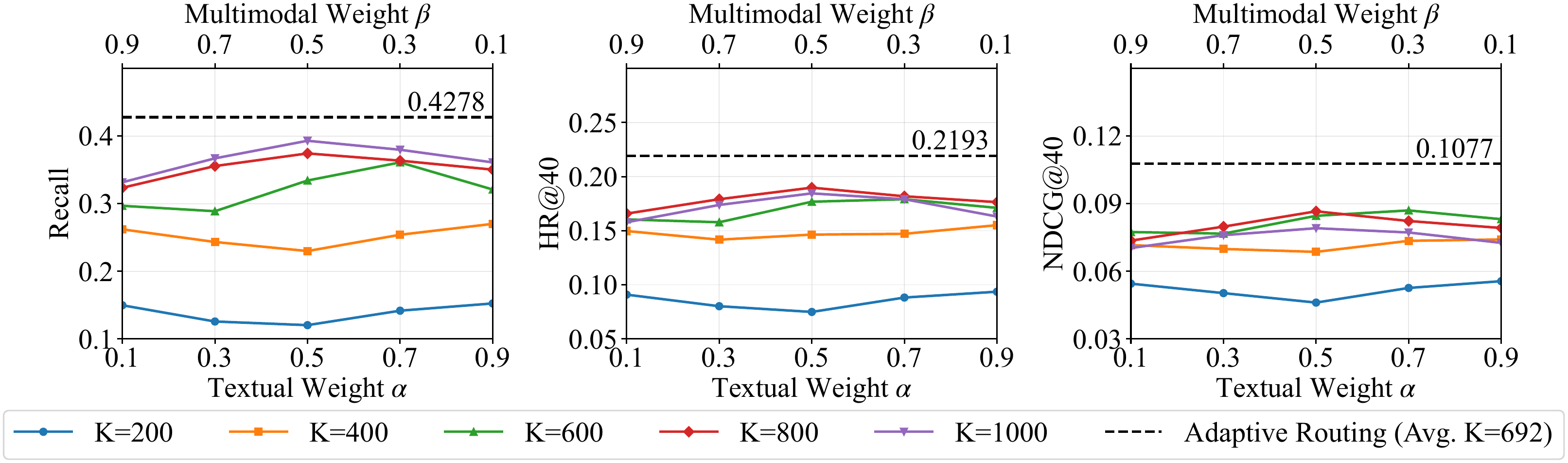}
    \caption{\textbf{Comparison of static/adaptive modality routing.} $K$ denotes total recall budget, while $\alpha$/$\beta$ denote text-based/multimodal recall weights, respectively. ``@40'' denotes HR/NDCG at top-40.}
    \label{fig:dis}
\vspace{-0.2in}
\end{figure*}

\noindent\textbf{Static and Adaptive Modality Routing.}
We analyze different recall budgets by varying the text-based recall weight and assigning the remaining weight to the multimodal branch. As shown in \autoref{fig:dis}, adaptive routing consistently outperforms static strategies, confirming the benefit of dynamically calibrating modality reliance and recall allocation for a specific query. Under fixed routing, when the recall budget is small, concentrating it on the more suitable modality can be more effective than evenly splitting it across branches, since each branch may otherwise receive too few candidates for reliable retrieval. As the budget increases, however, retrieving more candidates also increases the burden on downstream ranking, which may hinder final recommendation quality. These results suggest that fixed routing is difficult to generalize across recommendation requests, highlighting the need for adaptive per-query modality allocation.

\noindent\textbf{Limitations.}
Despite its superior performance, \ourmethod introduces additional computational overhead due to its agentic workflow. This reflects a common trade-off in agentic recommender systems, where more adaptive decision-making can improve recommendation quality at the cost of increased inference complexity. Therefore, making the inference process more efficient without compromising recommendation performance remains an important direction for future work.

\section{Conclusion}
We propose \ourmethod, to the best of our knowledge, one of the first systematic efforts to conceptually integrate adaptive modality routing into multimodal recommendation under an agentic paradigm. Since different query intents exhibit varying degrees of reliance on multimodal features and textual semantics, indiscriminate modal fusion may introduce noisy cues and impair recommendation quality. To address this challenge, \ourmethod introduces an innovative adaptive modality routing mechanism. Specifically, the agent generates pseudo-queries that align with the information granularity of the current query but target known items from the user's historical interactions. By simulating  verifiable proxy recall tasks with pseudo-queries to perform self-optimization, the framework agentically allocates the weights and budgets between the text-based and multimodal recall branches for the actual query at inference time. Comprehensive evaluations show that \ourmethod achieves highly competitive results against SOTA methods, providing evidence that adaptively routing modality reliance is a worthwhile exploration for improving multimodal recommender systems.

\newpage

\bibliographystyle{unsrt}
\bibliography{ref}

\begin{thebibliography}{10}

\bibitem{wang2026mllmrec}
Yu~Wang, Yonghui Yang, Le~Wu, Jiancan Wu, Hefei Xu, and Hui Lin.
\newblock Mllmrec-r1: Incentivizing reasoning capability in large language
  models for multimodal sequential recommendation.
\newblock {\em arXiv preprint arXiv:2603.06243}, 2026.

\bibitem{ye2025harnessing}
Yuyang Ye, Zhi Zheng, Yishan Shen, Tianshu Wang, Hengruo Zhang, Peijun Zhu,
  Runlong Yu, Kai Zhang, and Hui Xiong.
\newblock Harnessing multimodal large language models for multimodal sequential
  recommendation.
\newblock In {\em Proceedings of the AAAI Conference on Artificial
  Intelligence}, volume~39, pages 13069--13077, 2025.

\bibitem{fu2025vistawise}
Honghao Fu, Junlong Ren, Qi~Chai, Deheng Ye, Yujun Cai, and Hao Wang.
\newblock Vistawise: Building cost-effective agent with cross-modal knowledge
  graph for minecraft.
\newblock In {\em Proceedings of the 2025 Conference on Empirical Methods in
  Natural Language Processing}, pages 21895--21909, 2025.

\bibitem{tao2020mgat}
Zhulin Tao, Yinwei Wei, Xiang Wang, Xiangnan He, Xianglin Huang, and Tat-Seng
  Chua.
\newblock Mgat: Multimodal graph attention network for recommendation.
\newblock {\em Information Processing \& Management}, 2020.

\bibitem{zhai2025simple}
Penglong Zhai, Yifang Yuan, Fanyi Di, Jie Li, Yue Liu, Chen Li, Jie Huang,
  Sicong Wang, Yao Xu, and Xin Li.
\newblock A simple contrastive framework of item tokenization for generative
  recommendation.
\newblock {\em arXiv preprint arXiv:2506.16683}, 2025.

\bibitem{zhaimultimodal}
Jianyang Zhai, Zi-Feng Mai, Chang-Dong Wang, Feidiao Yang, Xiawu Zheng, Hui Li,
  and Yonghong Tian.
\newblock Multimodal quantitative language for generative recommendation.
\newblock In {\em The Thirteenth International Conference on Learning
  Representations}.

\bibitem{pomo2025recommender}
Claudio Pomo, Matteo Attimonelli, Danilo Danese, Fedelucio Narducci, and
  Tommaso Di~Noia.
\newblock Do recommender systems really leverage multimodal content? a
  comprehensive analysis on multimodal representations for recommendation.
\newblock In {\em Proceedings of the 34th ACM International Conference on
  Information and Knowledge Management}, pages 2377--2387, 2025.

\bibitem{liu2025onerec}
Zhanyu Liu, Shiyao Wang, Xingmei Wang, Rongzhou Zhang, Jiaxin Deng, Honghui
  Bao, Jinghao Zhang, Wuchao Li, Pengfei Zheng, Xiangyu Wu, et~al.
\newblock Onerec-think: In-text reasoning for generative recommendation.
\newblock {\em arXiv preprint arXiv:2510.11639}, 2025.

\bibitem{zhang2025improving}
Zhifang Zhang, Jiahan Zhang, Shengjie Zhou, Qi~Wei, Shuo He, Feng Liu, and Lei
  Feng.
\newblock Improving generalizability and undetectability for targeted
  adversarial attacks on multimodal pre-trained models.
\newblock {\em arXiv preprint arXiv:2509.19994}, 2025.

\bibitem{zhu2024bringing}
Xinliang Zhu, Sheng-Wei Huang, Han Ding, Jinyu Yang, Kelvin Chen, Tao Zhou, Tal
  Neiman, Ouye Xie, Son Tran, Benjamin Yao, et~al.
\newblock Bringing multimodality to amazon visual search system.
\newblock In {\em Proceedings of the 30th ACM SIGKDD Conference on Knowledge
  Discovery and Data Mining}, pages 6390--6399, 2024.

\bibitem{fu2025brainvis}
Honghao Fu, Hao Wang, Jing~Jih Chin, and Zhiqi Shen.
\newblock Brainvis: Exploring the bridge between brain and visual signals via
  image reconstruction.
\newblock In {\em ICASSP 2025-2025 IEEE International Conference on Acoustics,
  Speech and Signal Processing (ICASSP)}, pages 1--5. IEEE, 2025.

\bibitem{zhu2025query}
Hai Zhu, Yuankai Guo, Ronggang Dou, and Kai Liu.
\newblock Query-life: Query-aware language image fusion embedding for
  e-commerce relevance.
\newblock In {\em Proceedings of the 31st International Conference on
  Computational Linguistics: Industry Track}, pages 21--28, 2025.

\bibitem{lei2026unirec}
Zijie Lei, Tao Feng, Zhigang Hua, Yan Xie, Guanyu Lin, Shuang Yang, Ge~Liu, and
  Jiaxuan You.
\newblock Unirec: Unified multimodal encoding for llm-based recommendations.
\newblock {\em arXiv preprint arXiv:2601.19423}, 2026.

\bibitem{zhang2024beyond}
Dailing Zhang, Shiyu Hu, Xiaokun Feng, Xuchen Li, Meiqi Wu, Jing Zhang, and
  Kaiqi Huang.
\newblock Beyond accuracy: Tracking more like human via visual search.
\newblock {\em Advances in Neural Information Processing Systems},
  37:2629--2662, 2024.

\bibitem{wei2019mmgcn}
Yinwei Wei, Xiang Wang, Liqiang Nie, Xiangnan He, Richang Hong, and Tat-Seng
  Chua.
\newblock Mmgcn: Multi-modal graph convolution network for personalized
  recommendation of micro-video.
\newblock In {\em Proceedings of the 27th ACM international conference on
  multimedia}, 2019.

\bibitem{dang2025mllmrec}
Yuzhuo Dang, Xin Zhang, Zhiqiang Pan, Yuxiao Duan, Wanyu Chen, Fei Cai, and
  Honghui Chen.
\newblock Mllmrec: Exploring the potential of multimodal large language models
  in recommender systems.
\newblock {\em arXiv preprint arXiv:2508.15304}, 2025.

\bibitem{wu2025refineshot}
Hang Wu, Yujun Cai, Haonan Ge, Hongkai Chen, Ming-Hsuan Yang, and Yiwei Wang.
\newblock Refineshot: Rethinking cinematography understanding with foundational
  skill evaluation.
\newblock {\em arXiv preprint arXiv:2510.02423}, 2025.

\bibitem{mei2025a1}
Lingrui Mei, Shenghua Liu, Yiwei Wang, Baolong Bi, Yuyao Ge, Jun Wan, Yurong
  Wu, and Xueqi Cheng.
\newblock a1: Steep test-time scaling law via environment augmented generation.
\newblock {\em arXiv preprint arXiv:2504.14597}, 2025.

\bibitem{chen2025haif}
Jianing Chen, Zehao Li, Yujun Cai, Hao Jiang, Chengxuan Qian, Juyuan Kang,
  Shuqin Gao, Honglong Zhao, Tianlu Mao, and Yucheng Zhang.
\newblock Haif-gs: Hierarchical and induced flow-guided gaussian splatting for
  dynamic scene.
\newblock In {\em NeurIPS 2025}, 2025.

\bibitem{jiang2025recgpt}
Yangqin Jiang, Xubin Ren, Lianghao Xia, Da~Luo, Kangyi Lin, and Chao Huang.
\newblock Recgpt: A foundation model for sequential recommendation.
\newblock In {\em Proceedings of the 2025 Conference on Empirical Methods in
  Natural Language Processing}, pages 10140--10154, 2025.

\bibitem{zhu2024collaborative}
Yaochen Zhu, Liang Wu, Qi~Guo, Liangjie Hong, and Jundong Li.
\newblock Collaborative large language model for recommender systems.
\newblock In {\em Proceedings of the ACM Web Conference 2024}, pages
  3162--3172, 2024.

\bibitem{zheng2025universal}
Bowen Zheng, Hongyu Lu, Yu~Chen, Wayne~Xin Zhao, and Ji-Rong Wen.
\newblock Universal item tokenization for transferable generative
  recommendation.
\newblock {\em arXiv preprint arXiv:2504.04405}, 2025.

\bibitem{shenglanguage}
Leheng Sheng, An~Zhang, Yi~Zhang, Yuxin Chen, Xiang Wang, and Tat-Seng Chua.
\newblock Language representations can be what recommenders need: Findings and
  potentials.
\newblock In {\em The Thirteenth International Conference on Learning
  Representations}, 2025.

\bibitem{zhou2025openonerec}
Guorui Zhou, Honghui Bao, Jiaming Huang, Jiaxin Deng, Jinghao Zhang, Junda She,
  Kuo Cai, Lejian Ren, Lu~Ren, Qiang Luo, et~al.
\newblock Openonerec technical report.
\newblock {\em arXiv preprint arXiv:2512.24762}, 2025.

\bibitem{zhang2022gaze}
Dailing Zhang, Yinxiao Tian, Kaifeng Chen, and Kun Qian.
\newblock Gaze-directed visual grounding under object referring uncertainty.
\newblock In {\em 2022 41st Chinese Control Conference (CCC)}, pages
  3807--3811. IEEE, 2022.

\bibitem{zhu2024cost}
Jieming Zhu, Mengqun Jin, Qijiong Liu, Zexuan Qiu, Zhenhua Dong, and Xiu Li.
\newblock Cost: Contrastive quantization based semantic tokenization for
  generative recommendation.
\newblock In {\em Proceedings of the 18th ACM Conference on Recommender
  Systems}, 2024.

\bibitem{liu2025bridging}
Enze Liu, Bowen Zheng, Wayne~Xin Zhao, and Ji-Rong Wen.
\newblock Bridging textual-collaborative gap through semantic codes for
  sequential recommendation.
\newblock In {\em Proceedings of the 31st ACM SIGKDD Conference on Knowledge
  Discovery and Data Mining V. 2}, 2025.

\bibitem{shi2025llada}
Teng Shi, Chenglei Shen, Weijie Yu, Shen Nie, Chongxuan Li, Xiao Zhang, Ming
  He, Yan Han, and Jun Xu.
\newblock Llada-rec: Discrete diffusion for parallel semantic id generation in
  generative recommendation.
\newblock {\em arXiv preprint arXiv:2511.06254}, 2025.

\bibitem{wu2026camreasoner}
Hang Wu, Yujun Cai, Zehao Li, Haonan Ge, Bowen Sun, Junsong Yuan, and Yiwei
  Wang.
\newblock Camreasoner: Reinforcing camera movement understanding via structured
  spatial reasoning.
\newblock {\em arXiv preprint arXiv:2602.00181}, 2026.

\bibitem{ren2024representation}
Xubin Ren, Wei Wei, Lianghao Xia, Lixin Su, Suqi Cheng, Junfeng Wang, Dawei
  Yin, and Chao Huang.
\newblock Representation learning with large language models for
  recommendation.
\newblock In {\em Proceedings of the ACM web conference 2024}, pages
  3464--3475, 2024.

\bibitem{fu2024dp}
Honghao Fu, Yufei Wang, Wenhan Yang, Alex~C Kot, and Bihan Wen.
\newblock Dp-iqa: Utilizing diffusion prior for blind image quality assessment
  in the wild.
\newblock {\em arXiv preprint arXiv:2405.19996}, 2024.

\bibitem{hou2023learning}
Yupeng Hou, Zhankui He, Julian McAuley, and Wayne~Xin Zhao.
\newblock Learning vector-quantized item representation for transferable
  sequential recommenders.
\newblock In {\em Proceedings of the ACM Web Conference 2023}, pages
  1162--1171, 2023.

\bibitem{peng2025denoising}
Yingtao Peng, Chen Gao, Yu~Zhang, Tangpeng Dan, Xiaoyi Du, Hengliang Luo, Yong
  Li, and Xiaofeng Meng.
\newblock Denoising alignment with large language model for recommendation.
\newblock {\em ACM Transactions on Information Systems}, 43(2):1--35, 2025.

\bibitem{ren2025coherence}
Junlong Ren, Gangjian Zhang, Yu~Hu, Jian Shu, Hui Xiong, and Hao Wang.
\newblock Enhanced partially relevant video retrieval through inter-and
  intra-sample analysis with coherence prediction.
\newblock {\em arXiv preprint arXiv:2504.19637}, 2025.

\bibitem{fu2023sgcn}
Honghao Fu, Yilang Shen, Yuxuan Liu, Jingzhong Li, and Xiang Zhang.
\newblock Sgcn: a multi-order neighborhood feature fusion landform
  classification method based on superpixel and graph convolutional network.
\newblock {\em International Journal of Applied Earth Observation and
  Geoinformation}, 122:103441, 2023.

\bibitem{wang2024llmrg}
Yan Wang, Zhixuan Chu, Xin Ouyang, Simeng Wang, Hongyan Hao, Yue Shen, Jinjie
  Gu, Siqiao Xue, James Zhang, Qing Cui, et~al.
\newblock Llmrg: Improving recommendations through large language model
  reasoning graphs.
\newblock In {\em Proceedings of the AAAI conference on artificial
  intelligence}, volume~38, pages 19189--19196, 2024.

\bibitem{sun2025llmser}
Yuqi Sun, Qidong Liu, Haiping Zhu, and Feng Tian.
\newblock Llmser: Enhancing sequential recommendation via llm-based data
  augmentation.
\newblock {\em arXiv preprint arXiv:2503.12547}, 2025.

\bibitem{mei2025surveycontextengineeringlarge}
Lingrui Mei, Jiayu Yao, Yuyao Ge, Yiwei Wang, Baolong Bi, Yujun Cai, Jiazhi
  Liu, Mingyu Li, Zhong-Zhi Li, Duzhen Zhang, Chenlin Zhou, Jiayi Mao, Tianze
  Xia, Jiafeng Guo, and Shenghua Liu.
\newblock A survey of context engineering for large language models, 2025.

\bibitem{zhang2026unleashing}
Zhiyang Zhang, Junda She, Kuo Cai, Bo~Chen, Shiyao Wang, Xinchen Luo, Qiang
  Luo, Ruiming Tang, Han Li, Kun Gai, et~al.
\newblock Unleashing the native recommendation potential: Llm-based generative
  recommendation via structured term identifiers.
\newblock {\em arXiv preprint arXiv:2601.06798}, 2026.

\bibitem{hong2025llm}
Minjie Hong, Yan Xia, Zehan Wang, Jieming Zhu, Ye~Wang, Sihang Cai, Xiaoda
  Yang, Quanyu Dai, Zhenhua Dong, Zhimeng Zhang, et~al.
\newblock Llm-bs: Enhancing large language models for recommendation through
  exogenous behavior-semantics integration.
\newblock In {\em The Web Conference 2025}, 2025.

\bibitem{wang2025enhancing}
Zihan Wang, Jinghao Lin, Xiaocui Yang, Yongkang Liu, Shi Feng, Daling Wang, and
  Yifei Zhang.
\newblock Enhancing llm-based recommendation through semantic-aligned
  collaborative knowledge.
\newblock {\em arXiv preprint arXiv:2504.10107}, 2025.

\bibitem{ge2025focusingcontrastiveattentionenhancing}
Yuyao Ge, Shenghua Liu, Yiwei Wang, Lingrui Mei, Baolong Bi, Xuanshan Zhou,
  Jiayu Yao, Jiafeng Guo, and Xueqi Cheng.
\newblock Focusing by contrastive attention: Enhancing vlms' visual reasoning,
  2025.

\bibitem{hong2025eager}
Minjie Hong, Yan Xia, Zehan Wang, Jieming Zhu, Ye~Wang, Sihang Cai, Xiaoda
  Yang, Quanyu Dai, Zhenhua Dong, Zhimeng Zhang, et~al.
\newblock Eager-llm: Enhancing large language models as recommenders through
  exogenous behavior-semantic integration.
\newblock In {\em Proceedings of the ACM on Web Conference 2025}, pages
  2754--2762, 2025.

\bibitem{ye2026align3gr}
Wencai Ye, Mingjie Sun, Shuhang Chen, Wenjin Wu, and Peng Jiang.
\newblock Align$^3$gr: Unified multi-level alignment for llm-based generative
  recommendation.
\newblock In {\em Proceedings of the AAAI Conference on Artificial
  Intelligence}, volume~40, pages 16154--16162, 2026.

\bibitem{yuyao2022vision}
Ge~Yuyao, Cheng Yiting, Wang Jia, Zhou Hanlin, and Chen Lizhe.
\newblock Vision transformer based on knowledge distillation in tcm image
  classification.
\newblock In {\em 2022 IEEE 5th International Conference on Computer and
  Communication Engineering Technology (CCET)}, pages 120--125. IEEE, 2022.

\bibitem{yang2025agentdr}
Mingdai Yang, Nurendra Choudhary, Jiangshu Du, Edward~W Huang, Philip~S Yu,
  Karthik Subbian, and Danai Kourta.
\newblock Agentdr dynamic recommendation with implicit item-item relations via
  llm-based agents.
\newblock {\em arXiv preprint arXiv:2510.05598}, 2025.

\bibitem{zhang2024agentcf}
Junjie Zhang, Yupeng Hou, Ruobing Xie, Wenqi Sun, Julian McAuley, Wayne~Xin
  Zhao, Leyu Lin, and Ji-Rong Wen.
\newblock Agentcf: Collaborative learning with autonomous language agents for
  recommender systems.
\newblock In {\em Proceedings of the ACM Web Conference 2024}, pages
  3679--3689, 2024.

\bibitem{xia2026multi}
Yu~Xia, Sungchul Kim, Tong Yu, Ryan~A Rossi, and Julian McAuley.
\newblock Multi-agent collaborative filtering: Orchestrating users and items
  for agentic recommendations.
\newblock In {\em Proceedings of the ACM Web Conference 2026}, pages
  8649--8652, 2026.

\bibitem{wu2025personalized}
Zijing Wu, Leheng Sheng, Yuanlin Xia, Yi~Zhang, Yuxin Chen, and An~Zhang.
\newblock Personalized recommendation agents with self-consistency.
\newblock In {\em Companion Proceedings of the ACM on Web Conference 2025},
  pages 2978--2982, 2025.

\bibitem{fang2024multi}
Jiabao Fang, Shen Gao, Pengjie Ren, Xiuying Chen, Suzan Verberne, and Zhaochun
  Ren.
\newblock A multi-agent conversational recommender system.
\newblock {\em arXiv preprint arXiv:2402.01135}, 2024.

\bibitem{portugal2024agentic}
Ivens Da~Silva Portugal, Paulo Alencar, and Donald Cowan.
\newblock An agentic ai-based multi-agent framework for recommender systems.
\newblock In {\em 2024 IEEE International Conference on Big Data (BigData)},
  pages 5375--5382. IEEE, 2024.

\bibitem{kong2025think}
Xiaoyu Kong, Junguang Jiang, Bin Liu, Ziru Xu, Han Zhu, Jian Xu, Bo~Zheng,
  Jiancan Wu, and Xiang Wang.
\newblock Think before recommendation: Autonomous reasoning-enhanced
  recommender.
\newblock {\em arXiv preprint arXiv:2510.23077}, 2025.

\bibitem{chen2026shopx}
Jiacheng Chen, Tao Zhang, Manxi Lin, Dunxian Huang, Teng Shi, Honghao Fu,
  Mengyan Li, Xinming Zhang, Chenchi Zhang, Xuan Lu, Xiaotong Du, et~al.
\newblock Shopx: A foundation model for intent-to-item fulfillment in agentic
  shopping.
\newblock {\em arXiv preprint arXiv:2606.31693}, 2026.

\bibitem{shu2024rah}
Yubo Shu, Haonan Zhang, Hansu Gu, Peng Zhang, Tun Lu, Dongsheng Li, and Ning
  Gu.
\newblock Rah! recsys--assistant--human: A human-centered recommendation
  framework with llm agents.
\newblock {\em IEEE Transactions on Computational Social Systems},
  11(5):6759--6770, 2024.

\bibitem{yu2025intelligent}
Yuanqing Yu, Zhefan Wang, Chumeng Jiang, Xinyi Li, Jiayin Wang, and Min Zhang.
\newblock Intelligent agents with adaptive knowledge fusion for personalized
  recommendation.
\newblock In {\em Companion Proceedings of the ACM on Web Conference 2025},
  pages 2983--2987, 2025.

\bibitem{wang2025tunable}
Mingze Wang, Chongming Gao, Wenjie Wang, Yangyang Li, and Fuli Feng.
\newblock Tunable llm-based proactive recommendation agent.
\newblock In {\em Proceedings of the 63rd Annual Meeting of the Association for
  Computational Linguistics (Volume 1: Long Papers)}, pages 19262--19276, 2025.

\bibitem{xiao2025mmagentrec}
Xiaochen Xiao.
\newblock Mmagentrec, a personalized multi-modal recommendation agent with
  large language model.
\newblock {\em Scientific Reports}, 15(1):12062, 2025.

\bibitem{mei2026gateddifferentiableworkingmemory}
Lingrui Mei, Shenghua Liu, Yiwei Wang, Yuyao Ge, Baolong Bi, Jiayu Yao, Jun
  Wan, Ziling Yin, Jiafeng Guo, and Xueqi Cheng.
\newblock Gated differentiable working memory for long-context language
  modeling, 2026.

\bibitem{fu2026videostir}
Honghao Fu, Miao Xu, Yiwei Wang, Dailing Zhang, Jun Liu, and Yujun Cai.
\newblock Videostir: Understanding long videos via spatio-temporally structured
  and intent-aware rag.
\newblock In {\em Proceedings of the 64th Annual Meeting of the Association for
  Computational Linguistics (Volume 1: Long Papers)}, pages 35793--35805, 2026.

\bibitem{chen2025tokensnodessemanticguidedmotion}
Jianing Chen, Zehao Li, Yujun Cai, Hao Jiang, Shuqin Gao, Honglong Zhao, Tianlu
  Mao, and Yucheng Zhang.
\newblock From tokens to nodes: Semantic-guided motion control for dynamic 3d
  gaussian splatting, 2025.

\bibitem{ge2025framemind}
Haonan Ge, Yiwei Wang, Kai-Wei Chang, Hang Wu, and Yujun Cai.
\newblock Framemind: Frame-interleaved video reasoning via reinforcement
  learning.
\newblock {\em arXiv preprint arXiv:2509.24008}, 2025.

\bibitem{fu2025sdr}
Honghao Fu, Yongli Gu, Yidong Yan, Yilang Shen, Yiwen Wu, and Libo Sun.
\newblock Sdr-gain: A high real-time occluded pedestrian pose completion method
  for autonomous driving.
\newblock {\em IEEE Transactions on Intelligent Transportation Systems},
  27(1):972--982, 2025.

\bibitem{hu2023adaptive}
Hengchang Hu, Wei Guo, Yong Liu, and Min-Yen Kan.
\newblock Adaptive multi-modalities fusion in sequential recommendation
  systems.
\newblock In {\em Proceedings of the 32nd ACM international conference on
  information and knowledge management}, pages 843--853, 2023.

\bibitem{zhang2026test}
Zhifang Zhang, Bojun Yang, Shuo He, Weitong Chen, Wei~Emma Zhang, Olaf Maennel,
  Lei Feng, and Miao Xu.
\newblock Test-time attention purification for backdoored large vision language
  models.
\newblock In {\em CVPR}, 2026.

\bibitem{zhang2024defending}
Zhifang Zhang, Shuo He, Haobo Wang, Bingquan Shen, and Lei Feng.
\newblock Defending multimodal backdoored models by repulsive visual prompt
  tuning.
\newblock {\em NeurIPS}, 2025.

\bibitem{huang2026dmesr}
Mingyao Huang, Qidong Liu, Wenxuan Yang, Moranxin Wang, Yuqi Sun, Haiping Zhu,
  Feng Tian, and Yan Chen.
\newblock Dmesr: Dual-view mllm-based enhancing framework for multimodal
  sequential recommendation.
\newblock {\em arXiv preprint arXiv:2602.13715}, 2026.

\bibitem{ge2026should}
Haonan Ge, Yiwei Wang, Hang Wu, and Yujun Cai.
\newblock What should a streaming video model remember?
\newblock {\em arXiv preprint arXiv:2606.16353}, 2026.

\bibitem{wang2025leveraging}
Zhaoliang Wang, Baisong Liu, Weiming Huang, Tingting Hao, Huiqian Zhou, and
  Yuxin Guo.
\newblock Leveraging multimodal large language model for multimodal sequential
  recommendation.
\newblock {\em Scientific Reports}, 15(1):28960, 2025.

\bibitem{wang2025mmsrarec}
Haoyu Wang, Yitong Wang, and Jining Wang.
\newblock Mmsrarec: Summarization and retrieval augumented sequential
  recommendation based on multimodal large language model.
\newblock {\em arXiv preprint arXiv:2512.20916}, 2025.

\bibitem{wang2025multimodal}
Yu~Wang, Yonghui Yang, Le~Wu, Yi~Zhang, and Richang Hong.
\newblock Multimodal large language models with adaptive preference
  optimization for sequential recommendation.
\newblock {\em arXiv preprint arXiv:2511.18740}, 2025.

\bibitem{zhang2025tokenswap}
Zhifang Zhang, Qiqi Tao, Jiaqi Lv, Na~Zhao, Lei Feng, and Joey~Tianyi Zhou.
\newblock Tokenswap: Backdoor attack on the compositional understanding of
  large vision-language models.
\newblock {\em arXiv preprint arXiv:2509.24566}, 2025.

\bibitem{yu2026thought}
Haocheng Yu, Yaxiong Wu, Hao Wang, Wei Guo, Yong Liu, Yawen Li, Yuyang Ye,
  Junping Du, and Enhong Chen.
\newblock Thought-augmented planning for llm-powered interactive recommender
  agent.
\newblock In {\em Proceedings of the 32nd ACM SIGKDD Conference on Knowledge
  Discovery and Data Mining V. 1}, pages 1821--1832, 2026.

\bibitem{mcauley2015image}
Julian McAuley, Christopher Targett, Qinfeng Shi, and Anton Van Den~Hengel.
\newblock Image-based recommendations on styles and substitutes.
\newblock In {\em Proceedings of the 38th international ACM SIGIR conference on
  research and development in information retrieval}, pages 43--52, 2015.

\bibitem{he2016ups}
Ruining He and Julian McAuley.
\newblock Ups and downs: Modeling the visual evolution of fashion trends with
  one-class collaborative filtering.
\newblock In {\em proceedings of the 25th international conference on world
  wide web}, pages 507--517, 2016.

\bibitem{yang2025qwen3}
An~Yang, Anfeng Li, Baosong Yang, Beichen Zhang, Binyuan Hui, Bo~Zheng, Bowen
  Yu, Chang Gao, Chengen Huang, Chenxu Lv, et~al.
\newblock Qwen3 technical report.
\newblock {\em arXiv preprint arXiv:2505.09388}, 2025.

\bibitem{zhang2025qwen3}
Yanzhao Zhang, Mingxin Li, Dingkun Long, Xin Zhang, Huan Lin, Baosong Yang,
  Pengjun Xie, An~Yang, Dayiheng Liu, Junyang Lin, et~al.
\newblock Qwen3 embedding: Advancing text embedding and reranking through
  foundation models.
\newblock {\em arXiv preprint arXiv:2506.05176}, 2025.

\bibitem{li2026qwen3}
Mingxin Li, Yanzhao Zhang, Dingkun Long, Keqin Chen, Sibo Song, Shuai Bai,
  Zhibo Yang, Pengjun Xie, An~Yang, Dayiheng Liu, et~al.
\newblock Qwen3-vl-embedding and qwen3-vl-reranker: A unified framework for
  state-of-the-art multimodal retrieval and ranking.
\newblock {\em arXiv preprint arXiv:2601.04720}, 2026.

\bibitem{bai2025qwen3}
Shuai Bai, Yuxuan Cai, Ruizhe Chen, Keqin Chen, Xionghui Chen, Zesen Cheng,
  Lianghao Deng, Wei Ding, Chang Gao, Chunjiang Ge, et~al.
\newblock Qwen3-vl technical report.
\newblock {\em arXiv preprint arXiv:2511.21631}, 2025.

\bibitem{trotman2014improvements}
Andrew Trotman, Antti Puurula, and Blake Burgess.
\newblock Improvements to bm25 and language models examined.
\newblock In {\em Proceedings of the 19th Australasian Document Computing
  Symposium}, pages 58--65, 2014.

\bibitem{li2023making}
Chaofan Li, Zheng Liu, Shitao Xiao, and Yingxia Shao.
\newblock Making large language models a better foundation for dense retrieval,
  2023.

\bibitem{chen2024m3}
Jianlyu Chen, Shitao Xiao, Peitian Zhang, Kun Luo, Defu Lian, and Zheng Liu.
\newblock M3-embedding: Multi-linguality, multi-functionality,
  multi-granularity text embeddings through self-knowledge distillation.
\newblock In {\em Findings of the association for computational linguistics:
  ACL 2024}, pages 2318--2335, 2024.

\bibitem{yao2022react}
Shunyu Yao, Jeffrey Zhao, Dian Yu, Nan Du, Izhak Shafran, Karthik Narasimhan,
  and Yuan Cao.
\newblock React: Synergizing reasoning and acting in language models.
\newblock {\em arXiv preprint arXiv:2210.03629}, 2022.

\bibitem{shinn2023reflexion}
Noah Shinn, Federico Cassano, Ashwin Gopinath, Karthik Narasimhan, and Shunyu
  Yao.
\newblock Reflexion: Language agents with verbal reinforcement learning.
\newblock {\em Advances in neural information processing systems},
  36:8634--8652, 2023.

\bibitem{huang2025recommender}
Xu~Huang, Jianxun Lian, Yuxuan Lei, Jing Yao, Defu Lian, and Xing Xie.
\newblock Recommender ai agent: Integrating large language models for
  interactive recommendations.
\newblock {\em ACM Transactions on Information Systems}, 43(4):1--33, 2025.

\bibitem{wang2024macrec}
Zhefan Wang, Yuanqing Yu, Wendi Zheng, Weizhi Ma, and Min Zhang.
\newblock Macrec: A multi-agent collaboration framework for recommendation.
\newblock In {\em Proceedings of the 47th International ACM SIGIR Conference on
  Research and Development in Information Retrieval}, 2024.

\bibitem{hidasi2015session}
Bal{\'a}zs Hidasi, Alexandros Karatzoglou, Linas Baltrunas, and Domonkos Tikk.
\newblock Session-based recommendations with recurrent neural networks.
\newblock {\em arXiv preprint arXiv:1511.06939}, 2015.

\bibitem{kang2018self}
Wang-Cheng Kang and Julian McAuley.
\newblock Self-attentive sequential recommendation.
\newblock In {\em 2018 IEEE international conference on data mining (ICDM)},
  pages 197--206. IEEE, 2018.

\bibitem{singer2022sequential}
Uriel Singer, Haggai Roitman, Yotam Eshel, Alexander Nus, Ido Guy, Or~Levi,
  Idan Hasson, and Eliyahu Kiperwasser.
\newblock Sequential modeling with multiple attributes for watchlist
  recommendation in e-commerce.
\newblock In {\em Proceedings of the fifteenth ACM international conference on
  web search and data mining}, pages 937--946, 2022.

\bibitem{wu2025aligning}
Binrui Wu, Shisong Tang, Fan Li, Bing Han, Chang Meng, Jingyu Xiao, and Jiechao
  Gao.
\newblock Aligning and balancing id and multimodal representations for
  recommendation.
\newblock In {\em Proceedings of the 31st ACM SIGKDD Conference on Knowledge
  Discovery and Data Mining V. 2}, pages 5029--5038, 2025.

\bibitem{liao2024llara}
Jiayi Liao, Sihang Li, Zhengyi Yang, Jiancan Wu, Yancheng Yuan, Xiang Wang, and
  Xiangnan He.
\newblock Llara: Large language-recommendation assistant.
\newblock In {\em Proceedings of the 47th International ACM SIGIR Conference on
  Research and Development in Information Retrieval}, pages 1785--1795, 2024.

\bibitem{wei2024llmrec}
Wei Wei, Xubin Ren, Jiabin Tang, Qinyong Wang, Lixin Su, Suqi Cheng, Junfeng
  Wang, Dawei Yin, and Chao Huang.
\newblock Llmrec: Large language models with graph augmentation for
  recommendation.
\newblock In {\em Proceedings of the 17th ACM international conference on web
  search and data mining}, 2024.

\bibitem{bao2023tallrec}
Keqin Bao, Jizhi Zhang, Yang Zhang, Wenjie Wang, Fuli Feng, and Xiangnan He.
\newblock Tallrec: An effective and efficient tuning framework to align large
  language model with recommendation.
\newblock In {\em Proceedings of the 17th ACM conference on recommender
  systems}, pages 1007--1014, 2023.

\bibitem{gao2025sprec}
Chongming Gao, Ruijun Chen, Shuai Yuan, Kexin Huang, Yuanqing Yu, and Xiangnan
  He.
\newblock Sprec: Self-play to debias llm-based recommendation.
\newblock In {\em Proceedings of the ACM on Web Conference 2025}, pages
  5075--5084, 2025.

\bibitem{tschannen2025siglip}
Michael Tschannen, Alexey Gritsenko, Xiao Wang, Muhammad~Ferjad Naeem, Ibrahim
  Alabdulmohsin, Nikhil Parthasarathy, Talfan Evans, Lucas Beyer, Ye~Xia, Basil
  Mustafa, et~al.
\newblock Siglip 2: Multilingual vision-language encoders with improved
  semantic understanding, localization, and dense features.
\newblock {\em arXiv preprint arXiv:2502.14786}, 2025.

\end{thebibliography}

\newpage

\appendix

\section{Prompts}
\label{sec:prompt}

\begin{tcolorbox}[
  fonttitle = \small\bfseries,
  title=Multimodal Item Profiling Prompt,
  colframe=gray!2!black,
  colback=gray!2!white,
  boxrule=1pt,
  boxsep=0pt,
  left=5pt,
  right=5pt,
  fontupper=\setlength{\parskip}{2pt}\footnotesize, 
  halign title = flush center,
  breakable,
]

You are an expert e-commerce item profiler. Given product text and its main image, extract a fine-grained feature profile in STRICT JSON.

\textbf{Item text fields:} title: \{item.title\}, detail\_text: \{item.detail\_description\}, category\_hint: \{item.category\_hint or 'unknown'\}.

\textbf{Item profiling requirements:}

1) Type-first taxonomy:

   - Always output `item\_type` (required).
   
   - If hierarchical category is uncertain, keep only `item\_type` and leave `category\_path` empty.
   
   - If known, output `category\_path` as a list (e.g., ["Electronics", "Gaming", "Headset"]).
   
   - Also infer use\_case, target\_people, seasonality.
   
2) Textual attribute tags (fine-grained):

   - title keyword summary (must leverage title)
   
   - material/fabric composition
   
   - core features \& specs (size, capacity, weight, dimensions, compatibility, power, ingredients, etc.)
   
   - package/bundle information
   
   - quality \& durability claims
   
   - comfort/usability claims
   
   - price, price\_band inference (budget/mid/premium) and value\_for\_money signal
   
3) Visual attribute tags (from image):

   - dominant colors (+ optional hex-like names)
   
   - silhouette/shape/form factor
   
   - style keywords (minimalist, sporty, retro, luxury, kawaii, etc.)
   
   - texture/finish (matte/glossy/metallic/knit/grainy)
   
   - pattern/print/logo density
   
   - scene mood (formal, youthful, homey, professional, outdoor, gaming, etc.)
   
   - perceived quality level (low/medium/high with confidence)
   
4) Output ONLY one JSON object. No markdown.

\textbf{JSON schema:}

\{
  "item\_id": "\{item.item\_id\}",
  "title": "\{item.title\}",
  "taxonomy": \{
    "item\_type": "",
    "category\_path": [],
    "use\_case": [],
    "target\_people": [],
    "seasonality": ""
  \},
  "text\_attributes": \{...\},
  "visual\_attributes": \{...\}
\}

\end{tcolorbox}

\begin{tcolorbox}[
  fonttitle = \small\bfseries,
  title=Preference Modeling Prompt,
  colframe=gray!2!black,
  colback=gray!2!white,
  boxrule=1pt,
  boxsep=0pt,
  left=5pt,
  right=5pt,
  fontupper=\setlength{\parskip}{2pt}\footnotesize, 
  halign title = flush center,
  breakable,
]

You are a user preference modeling expert for an e-commerce recommendation system.

\textbf{Task:} Based on the user's relevant historical positive and negative behaviors, reason through the user's current preferences and shopping intent in chronological order.

\textbf{Requirements:}

1) Do not assume or reference query information that does not exist.

2) Clearly distinguish Strong\_Preference / Nice\_to\_Have / Dislike.

3) Preferences inferred from the positive sequence in the user history, especially positive behaviors, should be prioritized into Strong\_Preference or Nice\_to\_Have.

4) If the history contains analyzable visual information, such as visual\_tags or image-derived descriptions, Nice\_to\_Have must include conclusions about visual preferences. If there is no analyzable visual information, do not reference or fabricate non-existent visual information.

5) The history is already provided in ascending timestamp order: positive behaviors should be analyzed chronologically to infer evolving preferences and purchase trajectory; negative samples do not need to be reasoned about in chronological order.

6) Reasoning must combine contrastive evidence from both positive and negative behaviors in the history.

7) Strong\_Preference and Nice\_to\_Have do not have a fixed number of items; decide the number of items based on the strength of the evidence.

8) Every element in Strong\_Preference, Nice\_to\_Have, and Dislike must be a short natural-language preference phrase, such as “Nintendo Switch compatibility” or “wireless/Bluetooth connection”. Do not output JSON fragments, dictionaries, key-value pairs, or quoted structured snippets as elements.

9) Preconditions must take priority over general preferences. Do not output conclusions that conflict with the preconditions. The preconditions are as follows: \{preconditions\_prompt\}.

10) Output a strict JSON object with the following fields: Strong\_Preference, Nice\_to\_Have, Dislike, Reasoning\_Process.

Relevant historical records (JSON): \{json.dumps(history\_interactions\_with\_item\_profiles, ensure\_ascii=False)\}

\end{tcolorbox}

\begin{tcolorbox}[
  fonttitle = \small\bfseries,
  title=Preconditions Prompt,
  colframe=gray!2!black,
  colback=gray!2!white,
  boxrule=1pt,
  boxsep=0pt,
  left=5pt,
  right=5pt,
  fontupper=\setlength{\parskip}{2pt}\footnotesize, 
  halign title = flush center,
  breakable,
]

\textbf{Preconditions:}

A) First determine the product type of the user's current intent and the key preconditions before outputting preferences.

B) Any attribute that conflicts with the preconditions must be placed in Dislike, and conflicting items must not appear in Strong\_Preference or Nice\_to\_Have.

C) If a user query exists and the historical behaviors conflict with the current query, prioritize the preconditions from the current query; the history should only be used as supplemental evidence for style, budget, theme, or similar preferences.

D) Do not recommend cross-platform or incompatible products, such as PC games vs. PS/Xbox/Switch games, or iOS accessories vs. Android-only accessories.

E) For audience-sensitive categories, the target-audience preconditions must be checked, such as gender, age group, and sizing system.

F) For technical products, compatibility preconditions must be checked, such as system version, interface/protocol, power/voltage, and size specifications.

G) If there is insufficient information to confirm compatibility, explicitly include “avoid incompatible/platform-mismatched products” in Dislike, and explain the uncertainty in the Reasoning Process.

H) Before outputting, perform a consistency self-check: Strong\_Preference and Dislike must not contradict each other, and all Strong\_Preference items must satisfy the preconditions.

\end{tcolorbox}

\begin{tcolorbox}[
  fonttitle = \small\bfseries,
  title=Param Initialization Prompt,
  colframe=gray!2!black,
  colback=gray!2!white,
  boxrule=1pt,
  boxsep=0pt,
  left=5pt,
  right=5pt,
  fontupper=\setlength{\parskip}{2pt}\footnotesize, 
  halign title = flush center,
  breakable,
]

Please identify the common pattern across the similar users’ parameter results, and then generate the initialization parameters for a new user based on that common pattern. The input contains only three types of parameters:

- text\_weight: text weight

- vl\_weight: visual weight

- recall\_size: total recall budget

\textbf{Requirements:}

1) Output a consensus parameter triplet: text\_weight, vl\_weight, recall\_size. 

2) text\_weight + vl\_weight must equal 1, with a rounding tolerance of ±0.001.

3) Both text\_weight and vl\_weight must be within [0.30, 0.70]. As an initialization strategy, start from a balanced and conservative setting.

4) recall\_size must be within \{max\_recall\_size\} and must be an integer.

5) If these users show large disagreement, prioritize a robust middle value and do not let extreme values dominate.

6) Also include a reasoning field that briefly explains the common pattern among the similar users and why the final values were chosen.

7) Output only one JSON object, without any other explanatory text.

Input data: \{collaborative\_users\_params\}

\textbf{Strict output format:}

\{"text\_weight": "", "vl\_weight": "", "recall\_size": "", "reasoning": ""\}

\end{tcolorbox}

\begin{tcolorbox}[
  fonttitle = \small\bfseries,
  title=Pseudo Query Rewrite Prompt,
  colframe=gray!2!black,
  colback=gray!2!white,
  boxrule=1pt,
  boxsep=0pt,
  left=5pt,
  right=5pt,
  fontupper=\setlength{\parskip}{2pt}\footnotesize, 
  halign title = flush center,
  breakable,
]

Given an actual query and the profile of a history item, please generate a pseudo query that satisfies the following requirements:

1) The information granularity must be consistent with the actual query, neither broader nor more specific.

2) The semantic slot structure should be as consistent as possible, such as category, brand, specification, function, target audience, style, scenario, budget, and similar slots.

3) The specific content must point to the given history item and must be supported by its title, category, or attributes. Do not fabricate information.

4) Do not rewrite the history item into a full long description. Output only a search query that a user might type.

5) If the actual query contains hard constraints, such as brand, model, size, compatibility, or target audience, map them only when the history item provides corresponding evidence. If there is no evidence, keep the expression weakened at the same level of granularity and do not make things up.

6) Output only one line of pseudo query text, with no explanation.

7) If the content of the given historical item clearly mismatches the potential target of the actual query, **must** ignore the above conditions and directly output only "INVALID".

\textbf{Input:}

Actual query: \{actual\_query\}, History item information (JSON): \{history\_item\_json\}

Please output: <one-line pseudo query only>

\end{tcolorbox}

\begin{tcolorbox}[
  fonttitle = \small\bfseries,
  title=Modality Routing Prompt,
  colframe=gray!2!black,
  colback=gray!2!white,
  boxrule=1pt,
  boxsep=0pt,
  left=5pt,
  right=5pt,
  fontupper=\setlength{\parskip}{2pt}\footnotesize, 
  halign title = flush center,
  breakable,
]
You are a modality-routing controller for an adaptive multimodal recommendation system.

\textbf{Context:}
A history-aligned pseudo-query was generated for an item the user previously liked. Both text-based recall and vision-language recall were run on this pseudo-query. Since the target item is known, text\_rank and vl\_rank show which modality is more reliable for the current recommendation intent.

\textbf{Task:}
Update the routing parameters for the next iteration:
(1) text\_weight: reliance on text-based semantic recall.
(2) vl\_weight: reliance on visual or vision-language similarity recall.
(3) total\_recall: overall recall budget.

\textbf{Goal:}
Optimize text\_weight, vl\_weight, and total\_recall based on the pseudo-query, target item profile, text\_rank, vl\_rank, current routing state, memory, and constraints.

\textbf{Routing principles:}
(1) Lower rank is better.
(2) Strongly favor the modality that retrieves the target much better or retrieves it while the other misses.
(3) Prefer differentiated weights when evidence is clear. 
(5) Use moderate weights such as 0.65/0.35 or 0.7/0.3 when evidence is suggestive but not decisive.
(6) Use balanced weights only when evidence is genuinely mixed or weak.
(7) Must prioritize consistent memory trends over a single noisy observation.
(8) If current evidence conflicts with memory, update conservatively unless the new evidence is very strong.

\textbf{Text-dominant signals:}
Increase text\_weight when the evidence is mainly semantic, functional, or specification-driven, such as brand, model, compatibility, product type, function, material, size, capacity, wattage, voltage, flavor, scent, ingredient, quantity, title, or version. If text recall clearly beats VL recall, strongly favor text.

\textbf{VL-dominant signals:}
Increase vl\_weight when the evidence depends mainly on visual appearance, such as style, color, shape, silhouette, texture, pattern, decoration, aesthetic, design, look, packaging, or visual similarity. If VL recall clearly beats text recall, strongly favor VL.

\textbf{Mixed guidance:}
Use moderate or balanced weights when textual constraints and visual preferences are both important, ranks are close, wins alternate, or modalities provide complementary evidence. Prefer mild differentiation such as 0.6/0.4 or 0.7/0.3 over exactly 0.5/0.5 when there is any clear tendency.

\textbf{Recall-size guidance:}
(1) Increase total\_recall if both modalities miss, ranks are poor, or the query is broad or ambiguous.
(2) Decrease total\_recall if the dominant modality retrieves targets reliably at high ranks and the query is specific.
(3) Keep total\_recall stable when evidence is mixed or the previous value seems adequate.
(4) Do not increase recall only to compensate for a bad modality; reduce that modality's weight when confidence is high.

\textbf{Constraints:}
(1) text\_weight and vl\_weight must be floats in [0.0, 1.0].
(2) text\_weight + vl\_weight should be 1.0.
(3) total\_recall must be a positive integer.
(4) Obey input bounds, especially max recall size {max\_recall\_size} and min recall size {min\_recall\_size}.

\textbf{Output:}
Return strictly one valid JSON object only, with no markdown or extra text.

\textbf{JSON schema:}
\{
  "text\_weight": float,
  "vl\_weight": float,
  "total\_recall": int,
  "summary": string,
  "reasoning": string
\}

The reasoning field should briefly state which modality performed better, whether the query is text-driven, VL-driven, or mixed, how memory/rule\_hints affected the update, and why total\_recall changed or stayed stable.

\textbf{Input:}

\{
  "step": ...,
  "target\_item\_profile": \{history\_item\_profile\},
  "pseudo\_query": \{pseudo\_query\},
  "observation": \{
    "text\_rank": ...,
    "vl\_rank": ...
  \}
  "current\_state": ...,
  "memory": ...
  "constraints": \{
    "max\_recall\_size": ...,
    "min\_recall\_size": ...
  \}
\}

\end{tcolorbox}

\begin{tcolorbox}[
  fonttitle = \small\bfseries,
  title=Relevance Scoring Prompt,
  colframe=gray!2!black,
  colback=gray!2!white,
  boxrule=1pt,
  boxsep=0pt,
  left=5pt,
  right=5pt,
  fontupper=\setlength{\parskip}{2pt}\footnotesize, 
  halign title = flush center,
  breakable,
]

You are an expert in fine-ranking e-commerce recommendations. Please judge, from the user's perspective, how well the candidate product matches the user's current preferences, and give a score from 1 to 5.

\textbf{Scoring rules:}

[1] = Completely Irrelevant. Clearly violates Dislike, or obviously conflicts with the core intent of the user's query, and should not be recommended even if it satisfies some preferences.

[2] = Slightly Relevant. Only weakly related to the query, matching only a few superficial keywords or marginal attributes, and does not effectively satisfy the user's current core needs; or satisfies very few Strong\_Preference requirements.

[3] = Moderately Relevant, satisfies part of the query's core needs, or satisfies some Strong\_Preference / multiple Nice\_to\_Have requirements, but still has obvious gaps in product type, key attributes, or usage scenario.

[4] = Mostly Relevant, satisfies the query's core needs well, covers most Strong\_Preference requirements, and gains some bonus from Nice\_to\_Have, with no obvious conflicts.

[5] = Highly Relevant, fully satisfies the query's core needs and Strong\_Preference requirements, has no Dislike conflicts, and stands out in Nice\_to\_Have, Predicted\_Next\_Items, or the user's potential preferences.

\textbf{Input information:}

- User Query: \{query\}

- Strong\_Preference: \{strong\_pref\}

- Nice\_to\_Have: \{nice\_to\_have\}

- Dislike: \{dislike\}

- Candidate profile: \{item\_profile\}

\textbf{Output requirements:}
Output only one number: 1/2/3/4/5.
Do not output explanations, punctuation, spaces, or any other content.

\end{tcolorbox}

\section{Profile Formats}
\label{sec:formats}

\begin{tcolorbox}[
  fonttitle = \small\bfseries,
  title=Multimodal Item Profile Format Example,
  colframe=gray!2!black,
  colback=gray!2!white,
  boxrule=1pt,
  boxsep=0pt,
  left=5pt,
  right=5pt,
  fontupper=\setlength{\parskip}{2pt}\footnotesize, 
  halign title = flush center,
  breakable,
]

\{

"item\_id": ..., 

"taxonomy": \{"item\_type": ..., "category\_path": [...], "use\_case": [...], "target\_people": [...], "seasonality": ..., ...\}, 

"text\_attributes": \{"title\_keyword\_summary": [...],"material\_fabric\_composition": [...], "core\_features\_specs": \{"size": ..., "capacity": ..., "weight": ..., "dimensions": ..., "compatibility": ..., "power": ..., "ingredients": ..., "package\_bundle\_information": [...], "quality\_durability\_claims": [...], "comfort\_usability\_claims": [....], "price": ..., "price\_band\_inference": ..., "value\_for\_money\_signal": ..., ...\}, ...\}, 

"visual\_attributes": \{"dominant\_colors": [...], "silhouette\_shape\_form\_factor": ..., "style\_keywords": [...], "texture\_finish": ..., "pattern\_print\_logo\_density": ..., "scene\_mood": ..., "perceived\_quality\_level": ..., ...\}

\}

\end{tcolorbox}

\begin{tcolorbox}[
  fonttitle = \small\bfseries,
  title=User Preference Profile Format Example,
  colframe=gray!2!black,
  colback=gray!2!white,
  boxrule=1pt,
  boxsep=0pt,
  left=5pt,
  right=5pt,
  fontupper=\setlength{\parskip}{2pt}\footnotesize, 
  halign title = flush center,
  breakable,
]

\{

  "user\_id": ...,
  
  "preferences": \{
  "strong\_pref": [...],
    "nice\_to\_have": [...],
    "dislike": [...],
    "reasoning": ... \},
  
   "history\_meta": \{"historical\_item\_profiles": \{...\}, "behavior\_labels": \{...\}, "timestamps:" \{...\}\}

\}

\end{tcolorbox}

\newpage


\end{document}